%% file: main.tex
\documentclass[article]{aastex631}

\usepackage{hyperref}
\usepackage{color}
\usepackage{soul}
\usepackage{natbib}
\usepackage{hyperref}
\usepackage{amsmath,bm}
\usepackage{xspace}
\usepackage{graphicx}
\usepackage{enumitem}
\usepackage{amssymb}
\usepackage{xifthen}
\usepackage{hyperref}
\usepackage[normalem]{ulem}
\usepackage{comment}
\usepackage{amsmath}

\hypersetup{    
  colorlinks      = {true},
  linkcolor       = {blue},
  citecolor       = {blue},
  urlcolor        = {blue},
}

\graphicspath{{./}{figures/}}

\input{defs_custom.tex}

\newcommand{\ramses}{{\sc Ramses}}
\newcommand\agama{{\sc agama}}

\defcitealias{pla20a}{Planck (2020)}

\shorttitle{Thick stellar discs and baryon sloshing}
\shortauthors{Bland-Hawthorn et al.}

\begin{document}

\title{Turbulent gas-rich discs at high redshift: origin of thick stellar discs through 3D `baryon sloshing'}

\input{authors.tex}

\correspondingauthor{J. Bland-Hawthorn}
\email{jonathan.bland-hawthorn@sydney.edu.au}
 
\begin{abstract}
In response to recent observations from JWST and ALMA, we explore a new class of dynamically self-consistent models that mimics a plausible progenitor of the Milky Way over a wide range of disc gas fractions, $f_{\rm gas}$. The high gas surface densities encourage vigorous star formation, which in turn couples with the gas to drive turbulence. We show that this coupling through momentum recoil drives a random walk of the baryonic potential minimum with respect to total gravitational potential, $\Phi_{\rm tot}(R,\phi,z)$. The amplitude of the bulk motion depends on the feedback strength, which in turn is directly associated with $f_{\rm gas}$. At its most extreme, when gas is the sole contributor to the disc potential ($f_{\rm gas}=100$\%), the amplitude of the walk can reach up to $R\approx 5$ kpc within $\Phi_{\rm tot}$. The disc dominates over dark matter ($f_{\rm disc}\gtrsim 50$\%) within $R_s=2.2 R_{\rm disc}$, where $R_{\rm disc}$ is the exponential disc scale length. For a lower $f_{\rm disc}$ and/or $f_{\rm gas}$, the 3D sloshing amplitude and velocity are reduced. The combination of strong feedback and sloshing leads to the newly formed stars being dynamically heated and settling to a more spatially extended disc population. The 3D heating process is roughly isotropic but its effects are more noticeable in $\vert z\vert$ due to the initial dynamical coldness of the star-forming disc. Such a disc has enhanced [$\alpha$/Fe] stellar abundances and a vertical (but no radial) gradient in stellar age and metallicity, both consistent with the Milky Way's thick stellar disc.
\end{abstract}


\section{Introduction} \label{s:intro}

\subsection{The need for new simulations}
Most disc galaxies observed in the near field today went through an early phase ($z>1$) when the disc mass was dominated by gas \citep{tac20}. In order for a gas disc to support itself against collapse, high star formation rates must have been commonplace, consistent with the high gas surface densities \citep{GenzelEtAl2010}. The energetic feedback presumably coupled to the gas with moderate efficiency ($\sim 10$\%) and provided a natural source of (turbulent) pressure to sustain the gas-rich medium \citep{agertz2021,bla24}. This was an intensely energetic era that can be traced to $z\sim 10$ by our most powerful telescopes, including the Atacama Large Millimeter Array (ALMA) and the James Webb Space Telescope (JWST). Today, Milky Way analogues have a low gas fraction ($f_{\rm gas}\lesssim 10$\%), and so provide limited information on the gas processes that were active at those early times. Here, we seek to gain some physical insight about the nature of turbulent gas-rich discs at high redshift through controlled simulations made possible with the {\tt Nexus} framework \citep{tep24}.

\begin{figure}[]
    \centering
\includegraphics[width=0.95\textwidth]{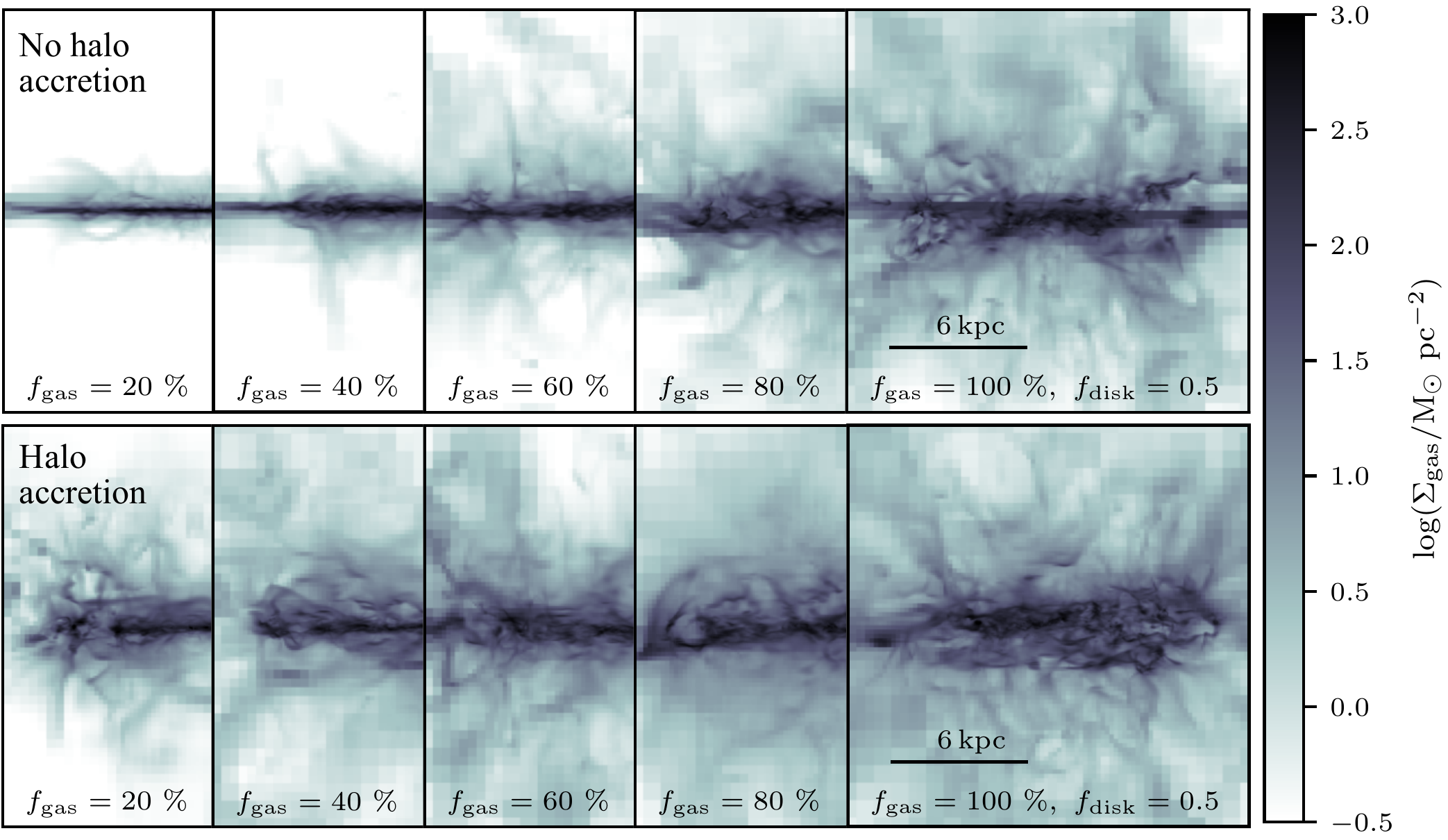}
\caption{Edge-on gas surface density maps taken from our early disc simulations at $t_{\rm o}=1\,{\rm Gyr}$. We adopt $f_{\rm disc}=0.5$ for the Milky Way progenitor at $z\sim 3$. The top panels are for five different initial gas fractions (as indicated) with {\it no} halo accretion; the bottom five are the matching simulations that include halo accretion from cooling hot gas. From left to right, for both rows, the galaxy's centre falls at the right-hand side (RHS) edge in the first four panels, but is centred in the last panel, where the spatial scale is indicated. The extent of the disc-halo interaction is a reflection of the disc-wide star formation shown in Fig.~\ref{f:sfr_dens}.
}
   \label{f:edgeon}
\end{figure}

Our new work is in response to mounting evidence for gas-dominant, star-forming discs in the redshift range $z\sim 1-8$ \citep{chap04,gen06h,forster06,shapiro08,swinbank12,hodge19,tsu21,riz22w,row24,rom24,tsu24}. There are clear cases where the gas entirely dominates the baryon disc fraction \citep[e.g.][]{rom24,row24,amvro25,hodge25}. To date, there has been limited work on realistic high-resolution simulations of isolated gas-rich galaxies sustained by underlying star formation and turbulence \citep[e.g.][]{van22}. 
Here we provide a sophisticated simulation framework ({\tt Nexus}; see \citealt{tep24}) to explore gas-dominant disc systems within the context of multi-component galaxy models (stellar disc and bulge, live dark matter halo, hot gaseous corona, multiphase ISM with star formation and metal production, etc.). Thus, it is now possible to explore disc evolution over the full range of $f_{\rm gas}=0-100\%$ in controlled experiments, as illustrated in Fig.~\ref{f:edgeon}.
(The parameters $f_{\rm disc}$ and $f_{\rm gas}$ are defined in Sec.~\ref{s:mod}.)

\begin{figure}[!htb]
    \centering
\includegraphics[width=0.49\textwidth]
{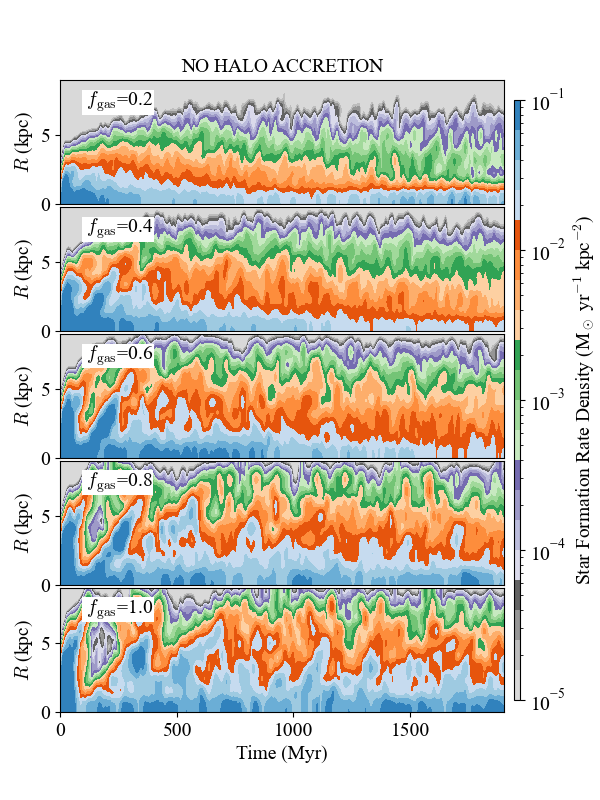}
\includegraphics[width=0.49\textwidth]
{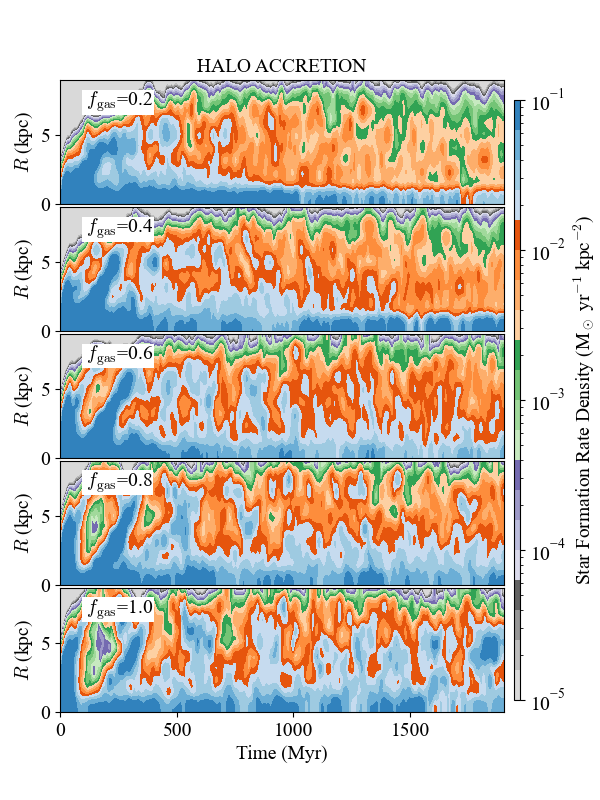}
\caption{
Periodograms drawn from our heavy disc ($f_{\rm disc}=0.5$) simulations that capture how the star formation rate (SFR) surface density (in units of M$_\odot$ yr$^{-1}$ kpc$^{-2}$) evolves as a function of radius and cosmic time: (Left) Without halo accretion; (Right) With halo accretion. The five rows correspond to different gas fractions, in order from the top, $f_{\rm gas}=20,40,60,80,100$\%. All models commence with a short-lived, disc-wide ``starburst'' phase that moves outwards for the first 300 Myr. These stochastic starbursts are defined by SFR surface densities that exceed the average rate by more than an order of magnitude (i.e. blue regions).
The higher $f_{\rm gas}$ models sustain a higher SFR surface density over a larger radial range for both models, as do the halo-accretion models in general when compared to the matching non-accreting model.
   }
   \label{f:sfr_dens}
\end{figure}

Our work has highlighted new processes in early gas-dominant discs that were missed in earlier (lower resolution) simulations. The {\tt Nexus} simulations reveal large-scale instabilities that arise in turbulent gas discs, triggered by pressure waves from stochastic star formation (see Fig.~\ref{f:sfr_dens}). These instabilities have important implications: as we now show, they drive disc asymmetries and baryon sloshing modes within the smooth gravitational potential.
Here, we focus on the latter and leave disc asymmetries to a companion paper \citep{bla25b}. 
We show how baryon sloshing provides us with an alternative heating mechanism leading to vertically-thickened stellar discs, which are observed in most disc galaxies \citep{yoa06,tsu25}. 

\subsection{A new era in disc research}

The formation of galactic discs has been one of the fastest developing areas of galactic astronomy in recent years. A host of observational results is revolutionising this field of research, including the discovery of dynamically cold galactic discs at very high redshift ($1<z<8$) with ALMA and JWST \citep[e.g.][]{2022ApJ...938L...2F,nel24,2023ApJ...955...94F,2023ApJ...946L..15K,2023ApJ...951L..46P}, the existence of already-mature discs containing bars \citep{costan23,rat25} and spiral structures \citep{2024ApJ...968L..15K}
as early as 2 Gyr after the Big Bang.
These discoveries concern objects that are expected to be massive, settling to discs earlier than Milky Way-like galaxies (disc downsizing, see \citealt{tsu25}), in some cases evolving into ellipticals in the present-day Universe. 
Exactly when Milky Way-like galaxies began to build their discs is uncertain \citep{mo98,san20a}. 
\cite{2023ApJ...951L..46P} present a galaxy at redshift 4.3 dominated by a stellar disc ($V_{\rm circ}/\sigma_{\rm LOS}\approx 5$) with baryon mass $M_{\star}\approx 2\times 10^9$ M$_{\odot}$,
a value in line with the estimated stellar mass of Milky Way (MW) precursors at such epochs \citep{2013ApJ...771L..35V}.

In our own Galaxy, the oldest and most metal-poor stars with disc kinematics are part of the thick disc and its metal-poor extension. The thick disc is traditionally defined in terms of its warm kinematics and its enhanced [$\alpha$/Fe] abundances compared to the solar value; this population is largely confined to $R<10$ kpc \citep{hay15n}.
Thick disc stars are thus expected to hold the key to understanding and constraining the physical mechanisms that govern how and when these dissipative structures formed. 
Unfortunately, the epoch and duration of formation of this population cannot be determined precisely from stellar ages, even with recent advances in asteroseismology \citep{sha19}. Some studies suggest thick disc formation lasted for 2--3 Gyr \citep{hayw13,2015A&A...578A..87S}, while others suggest a significantly faster formation, on the order of 1 Gyr \citep{2021A&A...645A..85M}.  In all cases, the bulk of the stellar mass of this population was in place 10~Gyr ago (at $z\approx 2$), with $M_{\star}$ in the range 20\% \citep{bla16} to 50\% \citep{2014ApJ...781L..31S} of the total present-day stellar mass of the disc.

The existence of such stellar discs at high redshifts requires that massive and early inflows of gas occurred in the inner kiloparsec regions of galaxies. Our understanding of the mechanism by which gas fed the inner regions of early galaxies has changed significantly in the last twenty years, with the identification of a cold mode of gas accretion in which gas is brought along a few dark matter filaments \citep{2003MNRAS.345..349B,2005MNRAS.363....2K,2006MNRAS.368....2D,2009MNRAS.395..160K}.
Although cold-mode accretion provides a mechanism by which the early disc building could be fed, simulating disc formation with cold ($V_{\rm circ}/\sigma_{\rm LOS}\gtrsim 3$) stellar kinematics -- such as the thick disc of the MW at epochs corresponding to $z>2$ -- continues to be very challenging \citep[][]{agertz09,2019MNRAS.490.3196P}. While more recent attempts appear promising \citep[e.g.][]{2022ApJ...928..106T,2024A&A...685A..72K}, {\it the formation of bimodal structures with thick and thin discs is not a natural outcome of these simulations.}

Historically, the first thick-disc formation scenarios focused on the response of a pre-existing, thin stellar component to various heating mechanisms: 
the accretion of stars during a merger \citep{1996A&A...305..125R,2003ApJ...597...21A}, the heating of the disc by an interaction or merger event \citep{1993ApJ...403...74Q,1996ApJ...460..121W,2008MNRAS.391.1806V,2011A&A...530A..10Q}, the diffusion of stellar orbits by massive star clusters \citep{2002MNRAS.330..707K,2011MNRAS.415.1280A} or by radial migration \citep{2011ApJ...737....8L}. Some studies also investigated the importance of a dissipative gaseous component in limiting the heating of a pre-existing stellar disc (see for example \citealt{2010MNRAS.403.1009M,2011MNRAS.415.3750M}).

After the first structures were identified in high-redshift discs \citep[e.g.][]{2005ApJ...627..632E,2008ApJ...687...59G,shapiro08}, the idea that massive gas clumps could provide a means of thickening the stellar component by diffusion of orbits was proposed by \cite{2009ApJ...707L...1B}. These authors carried out sticky particle simulations rather than more realistic hydrodynamic simulations.
As simulations of primordial discs improved \citep{dek09}, the proposal that thick discs could arise from stars forming in a thick turbulent ISM could finally be tested. Other work suggested that stars forming in a turbulent, kinematically hot ISM generate a thick stellar disc, which can be further heated by mergers or interactions with satellites \citep{2013MNRAS.436..625S,2017MNRAS.467.2430M,2021MNRAS.503.1815B,van22}. Here we challenge this claim by showing that the effect is not seen in higher-resolution {\tt Nexus} simulations.
Another scenario comes from \cite{men21} who, using cosmological simulations, find that it is the accumulation of the successive generations of stars forming in thin discs of different orientations that gradually produces a thick disc.

Previous scenarios have generally focused on providing a way to thicken a disc, or have assumed the stars are born thick.
However, the MW thick disc has important properties that were recently discovered, particularly from large spectroscopic surveys. First, the scale height of the thick disc is not constant, but increases from the most metal-rich and least $\alpha$-enhanced stars to the most metal-poor and most $\alpha$-enhanced (the so-called mono-abundance populations, see \citealt{2012ApJ...753..148B,mac17}). The $\alpha$-rich disc is radially compact, with a scale length of about 2~kpc which seems to be independent of the mono-abundant population \citep[see][their Fig. 5]{2012ApJ...753..148B}.
It is unclear whether the thick disc has a synchronized, monotonic, chemical evolution, with little radial dependence \citep{hayw13,2015A&A...579A...5H} or, conversely, an inside-out formation leading to a (steep) radial metallicity gradient \citep[e.g.][]{rat25}. The former case is expected to lead to a tight age-metallicity relation, as observed in several studies \citep{hayw13,xia22,cas24}, while the latter implies uncorrelated age-chemistry distributions 
\citep{2021A&A...645A..85M,2023A&A...678A.158A}.
All these features constitute further constraints for any model of early disc formation in the Milky Way. The aim of our paper is to revisit some of these ideas and to explore a new mechanism for disc thickening.

\smallskip
The structure of the paper is as follows.  In Section~\ref{s:mod}, we show how controlled simulations link to cosmological simulations, summarize the {\tt Nexus} models and provide links to the simulation movies.
In Section~\ref{s:proc}, we discuss the physical processes seen in the simulations, with a focus on the processes that lead to baryon sloshing.
In Section~\ref{s:pert}, we ask whether the rapid
growth of instabilities seen in the simulations are supported by theory. In Section~\ref{s:disc}, we discuss the implications of our new work and suggest future studies based on kinematic and morphological signatures. 
In Sec.~\ref{s:summ}, we place our new work in context and look forward to future improvements in the {\tt Nexus} framework.


\section{{\tt Nexus} simulations in brief} \label{s:mod}

\subsection{Controlled simulations in the context of cosmological N-body simulations}

In a seminal paper, \citet{mo98} introduced a simple theoretical framework describing how galaxy discs form from cooling hot gas within virialized dark matter halos, with their properties largely governed by angular momentum conservation (see also the updated treatment in \citet{ste20}). This work laid the foundation for semi-analytic models (SAMs) of galaxy formation. An unrecognized assumption in these models is that the circumgalactic medium (CGM) becomes structurally and dynamically stable (i.e., virialized) out to the virial radius at some early epoch. However, this assumption cannot hold in a strict sense, as the models also require that hot gas continues to cool within some inner radius.

Modern cosmological simulations diverge significantly in their predictions of CGM properties for Milky Way-like galaxies, mostly because of different implementations of star-formation driven galactic winds \citep[q.v.][]{ste21,wri24}. Controlled simulations are superior to cosmological simulations in tracing the evolution of galactic winds; the former include more realistic processes at much higher spatial resolution \citep[q.v.][]{sik25}.
Observationally, absorption-line studies of \HI\ and heavy elements along quasar sightlines suggest that the CGM began to settle into a more stable state by around $z\gtrsim 2$ \citep{chen12}, consistent with CGM models showing self-similar behavior since $z\sim 4$ \citep{pal14}. This agreement supports our working assumption for comparisons with controlled simulations, such as those in the {\tt VINTERGATAN} suite \citep{agertz2021}. Specifically, we adopt characteristic CGM properties from {\tt VINTERGATAN} progenitors at 
$z\approx 3$, a redshift by which the CGM appears largely stabilized (Rufo-Pastor et al., in prep.). In future work, we will compare the evolutionary trajectories produced by these complementary approaches to modelling galaxy evolution.

\subsection{Models and movies}
In earlier papers, we presented our new high-resolution (parsec scale) simulation framework ({\tt Nexus}), which is ideally suited to studying how turbulent discs evolve dynamically \citep{tep24}. These ecosystems manifest intermittent and long-lived behaviour, including disc-halo interaction, bars and spiral arms, and bulge formation \citep{bla24}. We were able to generate ``bar-like'' phenomena for the first time at all gas fractions, even in fully gas-dominated turbulent discs, including a remarkable ``radial shear flow'' with a characteristic signature that sets up before the central bulge is formed. As for the stellar discs, these bar-like signatures only arise in discs that dominate the local gravitational potential \citep{bla23}.

The choice of parameters for our progenitor analogues is described in \citet{bla24}.
This present work is focussed on objects with halo masses of order 10$^{11}$ M$_\odot$, a dynamical mass that is characteristic of the recent JWST discoveries at $z\sim 1-6$ \citep{guo23,leconte24,costan23}. This is the expected mass of a Milky Way progenitor at $z\approx 3$ \citep[Fig. 1 in][]{bla16}. In high-redshift galaxies, the observed high gas fraction and enhanced gas surface densities \citep{tac20} lead to elevated star-formation rates, and the output energy and momentum couple with the gas to drive the turbulence \citep{agertz09,jim23}. 
The adopted star/galaxy formation physics is presented in \citet{age13} and \citet{agertz2021}. The star formation activity feeds back sufficient energy and momentum to maintain the turbulent support. (We note, in passing, that turbulent gas-rich systems have also been explored where the energy injection arises from self gravity rather than star formation and feedback; see \citealt{fen23}.)

\begin{table*}
\begin{tabular}{rrccclcc}
\hline
$M_\star$ & $M_{\rm gas}$ & $R_{\rm disc}$ &  $f_{\rm disc}$   & $f_{\rm gas}$ & Label & High & Benchmark \\
($10^{8}$~\Msun)  & ($10^{8}$~\Msun) &  (ckpc) &  &  &  &  & \\
\hline
\hline
112 &  0.0  &  1.8  &  0.5    & 0.0 &  fd50\_fg00\_nac & 0 & 1 \\
88.8 &  22.2  &  1.8  &  0.5  & 0.2 &  fd50\_fg20\_nac & 2 & 1 \\
66.6 &  44.4  &  1.8  &  0.5  & 0.4 &  fd50\_fg40\_nac & 2 & 3 \\
44.4 &  66.6  &  1.8  &  0.5  & 0.6 &  fd50\_fg60\_nac & 2 & 1 \\
22.2 &  88.8  &  1.8  &  0.5  & 0.8 &  fd50\_fg80\_nac & 0 & 1 \\
0.0 &  112  &  1.8  &  0.5  & 1.0 &  fd50\_fg100\_nac  & 0 & 1 \\
\hline
88.8 &  22.2  &  1.8  &  0.5  & ((0.2)) &  fd50\_fg20\_ac & 0 & 1 \\
66.6 &  44.4  &  1.8  &  0.5  & ((0.4)) &  fd50\_fg40\_ac & 0 & 1 \\
44.4 &  66.6  &  1.8  &  0.5  & ((0.6)) &  fd50\_fg60\_ac & 0 & 1 \\
22.2 &  88.8  &  1.8  &  0.5  & ((0.8)) &  fd50\_fg80\_ac & 0 & 1 \\
0.0 &  112  &  1.8  &  0.5  & ((1.0)) &  fd50\_fg100\_ac & 0 & 1 \\
\hline
\hline\\
\end{tabular}
\caption{Overview of galaxy models of the initial parameters. The DM host halo properties ($M_{\rm halo} = 10^{11}$~\Msun; $R_{\rm vir} = 37$ ckpc; $r_{\rm s} = 9.2$ ckpc) are identical across models, and they are consistent with a MW-progenitor analogue at $z \approx 3$. Table columns are as follows: (1) pre-existing disc stellar mass (vertical scale height $=$ 250 pc for all models); (2) disc gas mass; (3) disc scale length; (4) disc-to-total mass ratio (Eq.~\ref{e:fd}); (5) Gas to total disc mass fraction at $t=0$ (Eq.~\ref{e:fdd}) $-$ the double brackets indicate accreting gas from a hot corona with a mass $M \approx 5.5\times10^9$~\Msun\ at $t = 0$; note that the hot halo mass is not accounted for in $f_{gas}$. (6) Model designation; (7,8) No.~of high and benchmark resolution models with different random seeds.}
\label{t:mod}
\end{table*}

Our first step is to consider an isolated galactic ecosystem in dynamical equilibrium, both with and without smooth accretion from the ambient hot corona.
All computations were carried out with the \ramses\ N-body/hydrodynamics code \citep[][]{tey02a} at benchmark ($N_{\rm lo}\sim 10^7$ elements) and high ($N_{\rm hi}\sim 10^8$ elements) resolution, including star formation and metal production (see Tab.~\ref{t:mod}); the number of effective gas ``cells'' is a factor of 10 higher in both cases.
In defining a Milky Way progenitor, we adopted a model with three key components: a live dark matter halo, a massive stellar/gaseous disc, and a hot coronal gas filling the live dark matter halo, which serves to supply the disc with a smooth flow of accreting gas after it cools.

For our Milky Way progenitor, we adopted halo parameters of $R_{\rm vir}\approx 40$ kpc and $\log M_{\rm vir}/{\rm M}_\odot \approx 11$. Two other key parameters were: (i) The disc mass fraction, $f_{\rm disc}$, which determines whether the disc baryons dominate the underlying dark matter halo,
\begin{equation}
\label{e:fd}
    f_{\rm disc}= \left(\frac{V_{\rm c, disc}(R_e)}{V_{\rm c, tot}(R_e)}\right)_{R_e=2.2 R_{\rm disc}}^2 \;.
\end{equation}
Here, $V_c(R)$ is the circular velocity at a radius $R$, $R_{\rm disc}$ is the exponential disc scale length, and $R_e=2.2 R_{\rm disc}$ is the traditional scale length adopted in studies of discs as it corresponds approximately to the ``turnover radius'' in the rotation curve\footnote{This radius contains about two-thirds of the disc's total mass.}; (ii) The gas mass fraction within the same radial scale:
\begin{equation}
\label{e:fdd}
    f_{\rm gas}=
    \left(\frac{M_{\rm disc, gas}}{M_{\rm disc}}\right)_{R_e=2.2 R_{\rm disc}}
\end{equation}
where $M_{\rm disc}$ is the total disc mass and $M_{\rm disc,gas}$ is the gas contribution. 

The values for all parameters are given in Table~\ref{t:mod}. The simulations are grouped into different categories $-$ benchmark resolution, high resolution, and repeated cases with different random seeds. There were 6 types of simulations distinguished by the different initial gas fractions ($f_{\rm gas}=0, 20, 40, 60, 80, 100\%$); these are run at benchmark resolution for a total of 2 Gyr because we are focussed on the high-redshift universe ($z\gtrsim 3$).


In all models, the gas fraction declines as more of the mass is locked up in stars \citep[see Fig. 3 in][]{bla24}. For models with $f_{\rm gas}<100\%$, we included a pre-existing (`old') stellar disc such that $f_{\rm old}=1-f_{\rm gas}$. For models with $f_{\rm gas} \approx 100\%$, we include a tiny pre-existing stellar population ($f_{\rm old}\approx 1\%$) as a useful tracer of the galaxy's gas-driven evolution. The total mass of stars produced in 2 Gyr from low to high gas fraction spans $2.6-6.2 \times 10^9$ M$_\odot$ for the accreting halo models, and $0.6-4.0 \times 10^9$ M$_\odot$ for the non-accreting models.
The baryon mass was preserved in both the accreting and non-accreting halo models. In the former case, the total baryon mass of the disc increases with time. Note that the initial scale length of the disc (both gas and stars) was roughly maintained at $R_{\rm disc}\approx 1.8$~kpc across models.
The above parameters are broadly consistent with arguably the best Milky Way progenitor analogue to date, the object CEERS-2112 at a photometric redshift of $z\approx 3.0$ \citep{costan23}.


In Table~\ref{t:mod}, the filename convention is {\tt fdXX\_fgYY\_nac} for ``no halo accretion'' models, and {\tt fdXX\_fgYY\_ac} for ``halo accretion'' models. Here, {\tt XX} is the disc mass fraction percentage and {\tt YY} is the gas fraction percentage. If the simulation has multiple versions generated by different random seeds, the file is referred to as {\tt fdXX\_fgYY\_nac.ZZ}, where {\tt ZZ} is the version number.
At our website,\footnote{ \href{http://www.physics.usyd.edu.au/turbo\_disks/}{http://www.physics.usyd.edu.au/turbo\_discs/}} we provide a series of movies (animations) that show the evolution of each of our model galaxies. There are two types of animations: 1) `on-the-fly' animations; and 2) `post-processed' animations.
The `on-the-fly' animations were created at simulation runtime; these have a high time resolution ($\delta t \approx 1$~Myr), and they show the evolution of the newly formed stars, of the gas density, and of the gas temperature on a face-on projection. The `post-processed' animations have been created from the simulation outputs; they have a lower time resolution ($\delta t \approx 10$~Myr), and they show the evolution of the surface density of all disk components (gas, pre-existing stars, and - if available - newly formed stars, gas) along three orthogonal projections.

\subsection{Reference frames}

An idealised, multi-component galaxy system is made up of several distinct contributions to the gravitational potential. Since the initial galaxy is set up as an equilibrium figure using \agama\ \citep{vas19a}, this inertial reference frame conserves momentum in all components, for both gas, stars and dark matter. In N-body simulations, the different mass components can become separated, depending on the nature and dynamics of the dark matter halo. In one of their examples, \citet{Joshi24} plot the separate trajectories for the centre of mass of the halo, bulge and disc. Once the baryonic component starts to slosh, since each subsystem has its own internal structure, the restoring force and the drag imposed by dynamical friction can be very different.


As the gas fraction increases, the system experiences stronger gravitational potential fluctuations, primarily driven by large-scale turbulent energy sustained by ongoing star formation. In particular, momentum recoil from sporadic, galaxy-wide starbursts adds significantly to these background fluctuations. As we demonstrate, the cumulative effect is a pronounced sloshing of the baryonic matter within the overall gravitational potential well. We find it useful and revealing to define two distinct centers that are tied to two different reference frames, with the first being defined by the minimum of the baryonic gravitational potential:
\begin{equation}
\nabla\Phi_{\rm bary}(R,\phi,z) \approx 0, \;\;\;\;\;\; \nabla^2\Phi_{\rm bary}(R,\phi,z) > 0
\label{e:bpot}
\end{equation}
with the Hessian matrix of partial second derivatives for $\Phi_{\rm bary}$ being positive definite. This baryonic centre ($x_{\rm bary}$, $y_{\rm bary}, z_{\rm bary}$) is useful because it happens to coincide with the peak of the light (surface density) distribution throughout. This is most often how observers perceive the centre of a galaxy.

The second reference frame defines the minimum of the total potential:
\begin{equation}
\nabla\Phi_{\rm tot}(R^\prime,\phi^\prime,z^\prime) \approx 0, \;\;\;\;\;\; \nabla^2\Phi_{\rm tot}(R^\prime,\phi^\prime,z^\prime) > 0
\label{e:tpot}
\end{equation}
with all partial second derivatives for $\Phi_{\rm tot}$ being positive definite. This centre of the total potential ($x_{\rm tot}$, $y_{\rm tot}$, $z_{\rm tot}$) is not traced by light, typically, in sloshing discs, only inferred from the disc kinematics. 
In practice, the minima defined in Eqs.~\ref{e:bpot} and \ref{e:tpot} are identified after convolving the gravitational field at each timestep with a window function defined by 50 nearest-neighbour particles. Given that the central baryons dominate over dark matter, we anticipate that both minima have similar sloshing amplitudes with respect to an inertial frame.


\begin{figure}[!htb]
\includegraphics[width=\textwidth]{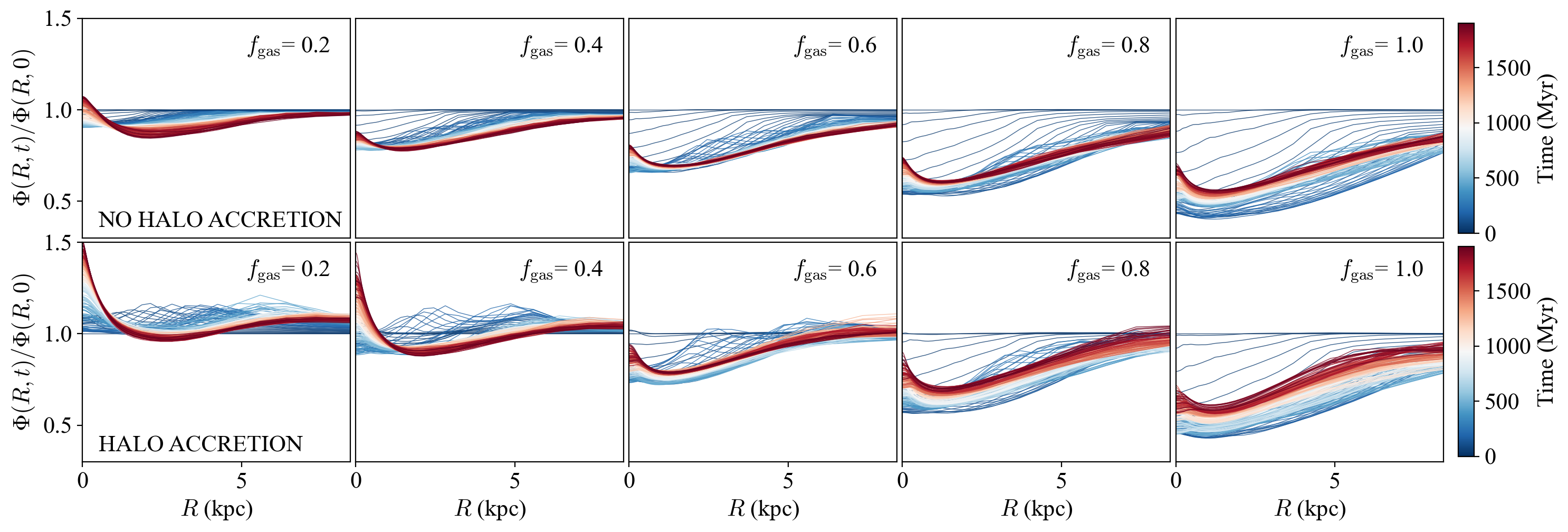}
    \caption{The evolution of the azimuthally averaged, radial profile of the disc potential $\Phi_{\rm bary}(R,t)$ with cosmic time, normalised to the starting potential $\Phi_{\rm bary}(R,0)$ encoded in colour for (top) no halo accretion, (bottom) halo accretion. Blue tracks are early in cosmic time and red tracks are later times, as indicated. For $f_{\rm gas} < 0.5$, the disc potential is fairly constant but, above this limit, the loss of gas mass in circulating winds leads to a substantial weakening of the disc potential. Note that the vertical axis is a ratio of gravitational potentials, such that a curve moving downwards reflects a weaker potential.
    }
   \label{f:pots0}
\end{figure}

\section{The origins of sloshing: simulations}
\label{s:proc}

\subsection{Unbinding the disc} \label{s:unbind}

In Fig.~\ref{f:pots0}, we show how the radial profile of the simulated disc's gravitational potential $\Phi_{\rm bary}(R,t)$ evolves with time $t$ as a function of the disc gas fraction $f_{\rm gas}$. For the `no halo accretion' models, there is only a 15$-$20\% change for $f_{\rm gas} < 0.5$.  In the halo accretion models, there is substantial accretion to the centre at low $f_{\rm gas}$, but the overall radial profile is relatively unchanged for $f_{\rm gas} < 0.5$. At higher $f_{\rm gas}$, the gravitational potential decreases rapidly over the first 500 Myr for both accretion cases. Up to half of the gravitational potential is lost due to gas launched into the halo by strong disc-wide winds. (The same process has been discussed in the context of dwarf galaxies by \citet{zha02}.)
Over the next 1.5 Gyr, there is a slow re-accretion to the inner disc, but most of it is recirculated to the outer disc through recycling in the galactic halo. The slow spin of the halo ensures the gas acquires angular momentum. We will explore the physics of this process in a later paper.

\begin{figure}[!htb]
\centering
\includegraphics[width=0.6\textwidth]{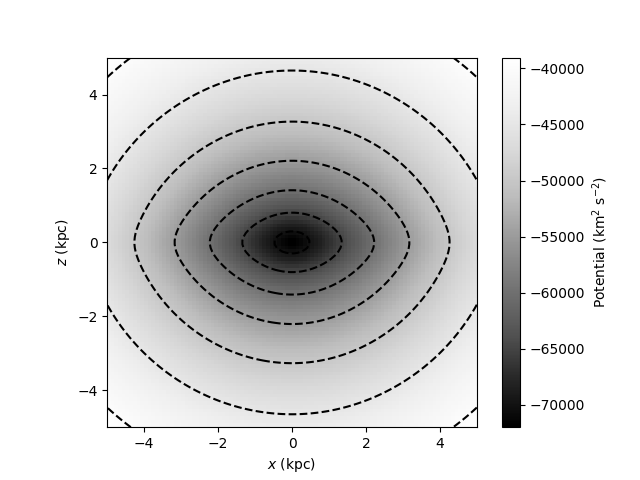}
    \caption{The total gravitational potential for a {\tt Nexus} galaxy simulation with $f_{\rm disc}=0.5$ and $f_{\rm gas}=0$. The equipotential surfaces are shown for the inner 10 kpc (comoving) diameter.
    }
   \label{f:totalpot}
\end{figure}

A great deal of energy is needed to `unbind' the disc in this way, but the feedback requirements through active star formation are consistent with high-redshift ALMA and JWST observations. We can obtain a rough estimate from the following formula for the binding energy:
\begin{equation}
    E_{\rm bind} = -\frac{3}{5}(1+\frac{q^2}{2})\frac{G M_{\rm disc}^2}{R_{\rm disc}}
\end{equation}
assuming the disc dominates the local gravitational potential. The factor $q$ is the degree of flattening of the oblate spheroid along the polar axis, or about $q\approx 0.7$ from Fig.~\ref{f:totalpot}. 
For convenience, we modify our estimate in terms of the disc's local surface density $\Sigma(R)= \Sigma_0 \exp(-R/R_{\rm disc})$ such that
\begin{equation}
    E_{\rm bind} \approx -3 \pi^2 G \Sigma_0^2 R_{\rm disc}^3 .
\end{equation}
In practice, this is an upper limit because the unbound material does not need to reach infinity.

Alternatively, the timescale $\Delta\tau$ needed to partially unbind (loosen) the disc is essentially the time it takes to lose half of its mass or, equivalently, its surface density \citep[e.g.][]{hil80}. Thus
\begin{equation}
    \Sigma(R,t) = \Sigma(R) - \dot{\Sigma}\; \Delta \tau
\end{equation}
where $\dot{\Sigma}=\epsilon \dot{\mu_*}$, for which $\dot{\mu}_\star$ is the SFR surface density, and
where $\epsilon$ ($=0.1$) is the coupling efficiency of the feedback energy and radiation 
with the gas \citep[][see their Sec. 5.2]{ego23}.
Here, $\Delta\tau$ is the duration of the starburst,
\begin{equation}
    \Delta \tau = \frac{\Sigma(R)}{2\epsilon \dot{\mu_*}(R)} .
    \label{e:losemass}
\end{equation}
The SFR surface density $\dot{\mu}_\star$ is the quantity shown in Fig.~\ref{f:sfr_dens} averaged over annular bins at each radius.
Therefore, after rearranging Eq.~\ref{e:losemass} and inserting the simulated surface density profile $\Sigma(R)$,
\begin{equation}
    \dot{\mu}_\star(R) = 27 \left(\frac{\Sigma_0\; e^{-R/R_{\rm disc}} }{5.45\times 10^8 \; \rm M_\odot\; kpc^{-2}}\right) \left(\frac{0.1}{\epsilon}\right)\left(\frac{100\;\rm Myr}{\Delta\tau} \right)\;\;\rm M_\odot\; kpc^{-2}\;yr^{-1} .
\end{equation}
This is about $\dot{\mu}_\star=0.3\pm 0.2$ M$_\odot$ kpc$^{-2}$ yr$^{-1}$ at $R\approx 8$ kpc where the disc is more reactive and easily unbound, consistent with Fig.~\ref{f:sfr_dens}, where sustained bursts over 100 Myr are a regular occurrence, hence the choice of normalization for $\Delta\tau$ in the above equation.
The simulated bursts can be much higher but these are occurring in compact regions and on shorter timescales than $\Delta\tau$. The inferred time-averaged rates are sufficiently impulsive to (partially) unbind the disc, particularly at large galactocentric radius. Here, we have taken the $R_{\rm disc}$ ($=1.7\pm 0.3$ kpc) and $\Sigma_0$ normalizations from our simulated models (see below). 
At these early times, the most powerful starbursts create of order $3\times 10^9$ stars in a $\Delta \tau$ interval across the inner kiloparsec of the disc.

We see that the required SFR surface density is achieved for all halo accretion models in the first 500 Myr. But at low $f_{\rm gas}$, the disc potential is dominated by the existing stellar populations and these remain unmoved by feedback processes. It is only at high values of $f_{\rm gas}$ that the disc becomes marginally unbound.\footnote{There is a strong similarity here with the dissolution of open star clusters \citep{bla10}. At the cluster's birth, the hot young stars drive the gas away and, if the surviving gas dominates the cluster's gravitational potential, the cluster becomes unbound. This is the fate of most open clusters since star formation efficiencies are consistently low in most environments, as observed in our models.}
The consequences of the unbinding are discussed in the next section.

\subsection{Stochastic fluctuations \& increasing entropy} \label{s:fluc}

In Fig.~\ref{f:pots0}, the unbinding of the disc at high $f_{\rm gas}$ appears to show a smooth evolution but these profiles are radially averaged. Locally, the unbinding is highly impulsive. Because the disc mass is lost through radiation pressure and entrainment in winds, the stellar disc becomes inflated vertically and radially. Stars are now oscillating vertically and radially in a weakened potential while retaining the same orbital energy. As we see later, this is not a smooth outward progression over hundreds of Myr. The stochastic nature of the starburst-driven fluctuations is crucial to the inflated disc remaining in that state.

In fact, we note that for high $f_{\rm gas}$ models, the disc reaccretes at later times. One might imagine that the process is reversible if the gravitational potential evolved smoothly, allowing the orbital properties of stars to evolve adiabatically to their former state. But this can never happen. The inflated stellar orbits are progressively more radial during expansion \citep[e.g.][]{yu21}. The randomized radial phases contribute to the overall increase in entropy and this is an irreversible process.

\begin{figure}[htb]
\centering
\includegraphics[width=0.45\textwidth]{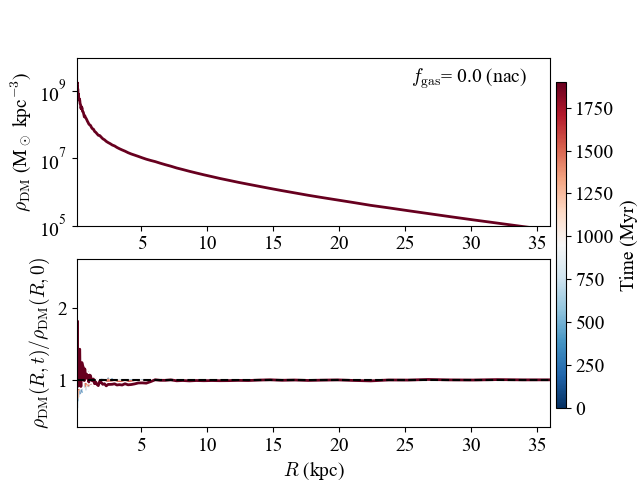}
\includegraphics[width=0.45\textwidth]{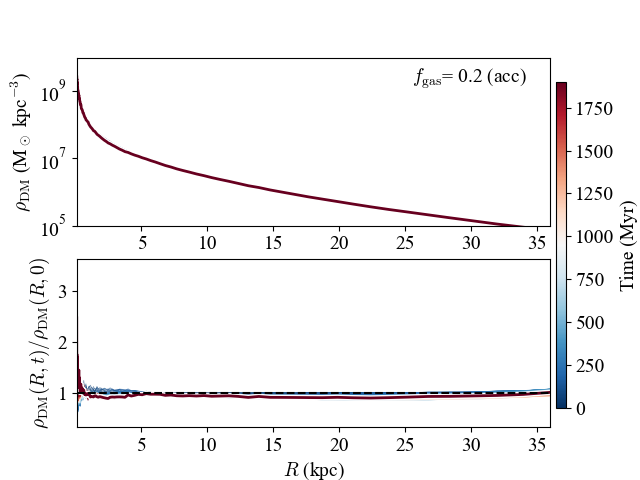}
\includegraphics[width=0.45\textwidth]{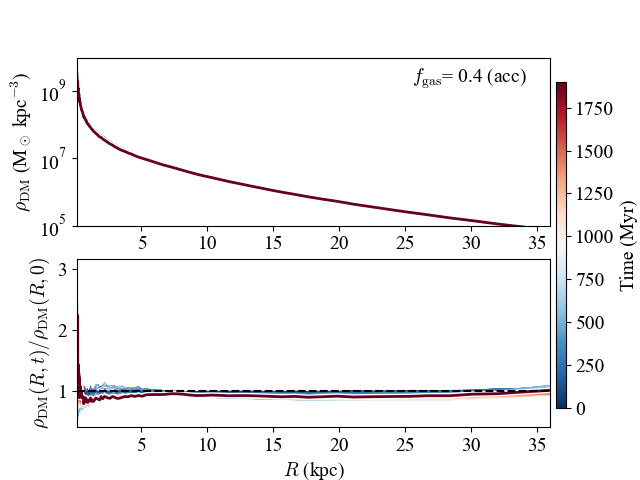}
\includegraphics[width=0.45\textwidth]{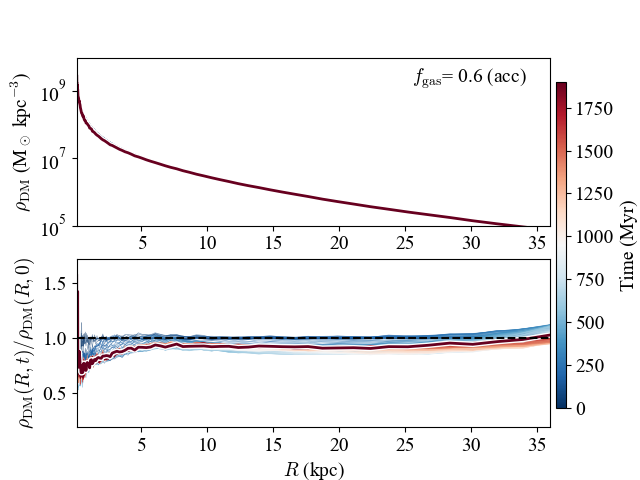}
\includegraphics[width=0.45\textwidth]{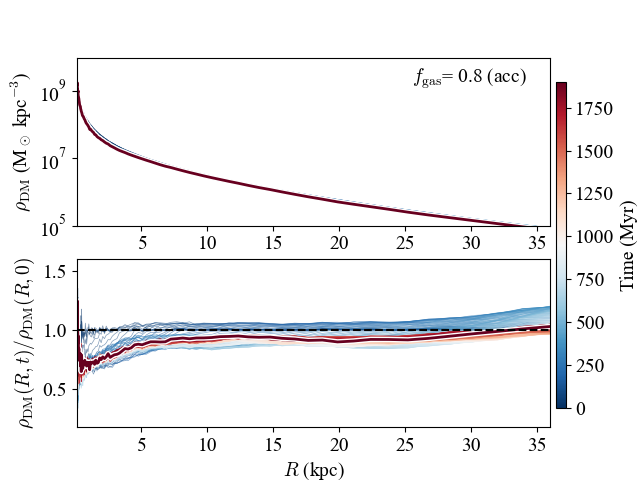}
\includegraphics[width=0.45\textwidth]{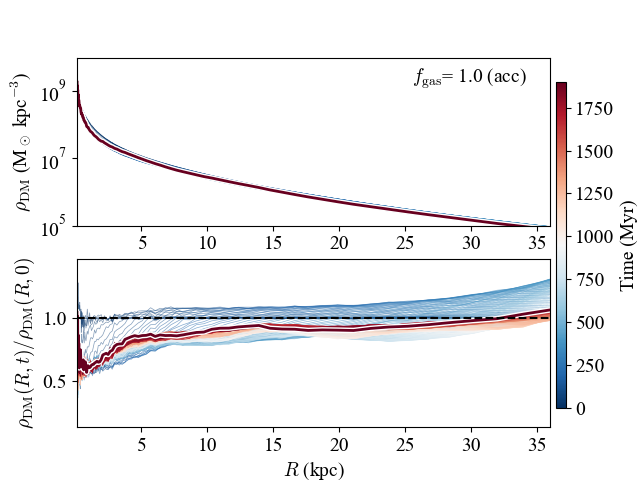}

    \caption{The upper figure in each panel shows evolution of the dark-matter density profile as a function of radius $R$ for the halo accretion models. The lower figure in each panel presents the ratio of the post-starburst to original density profiles. At the highest gas fractions, the central NFW cusp is suppressed by a factor of two, but not destroyed. In all plots, the colour coding indicates the time evolution of the density profile, as indicated, with the reddest profile indicating the end of the simulation.
    }
   \label{f:cusp}
\end{figure}

There is an interesting precedent for the disc inflation process in the context of dwarf galaxies \citep{zha02}.  A more extensive and related literature focusses on unbinding the NFW cusp predicted in dark-matter only simulations, but never observed. For example,
\citet{pon12n} show how fluctuations in the central gravitational potential $\Phi_0$ irreversibly transfer energy into collisionless particles, thus removing any dark matter `cusp' \citep{nav96a}. They consider blowout-recollapse cycles driven by a central starburst \citep[see also][]{ogi11,nip15}. By analogy, consider a star that is oscillating vertically with orbital frequency $\nu$ under simple harmonic motion
\begin{equation}
    z(t) = a(t) \cos[\nu(R) t + \psi(t)]
\end{equation}
where the orbital amplitude $a$ and phase $\psi$ that can undergo slow or rapid changes. The orbital frequency can be determined from
\begin{equation}
    \nu^2(R) \approx 4\pi G \Sigma(R)/ z_{\rm disc}
    \label{e:zfreq}
\end{equation}
where $z_{\rm disc}$ is the disc's vertical scale height. (The formal definition is given below.)
The key point is that a prompt change (i.e. faster than the orbital period) of $\nu$ at a fixed radius due to a sudden blowout and recollapse is now associated with a different amplitude $a_1$ and phase $\psi_1$ along the star's orbit compared to the original amplitude $a_0$ and phase $\psi_0$ \citep[cf.][]{yu21}. According to \citet{pon12n}, the change in energy has an explicit dependence on the orbital phase, such that
\begin{equation}
    \Delta E = \left[\left(\frac{\nu_1}{\nu_0}\right)^2 - 1 \right]\sin^2 \psi_0
\end{equation}
where $\nu_0$ and $\nu_1$ are the vertical frequencies immediately before and after blow-out or recollapse. This tends to extract energy from a star's orbit at the blow-out phase and return energy to the orbit at recollapse, but the overall effect is an energy gain due to the randomized orbital phases.\footnote{The analogy here is pushing a child on a swing. If the applied force is introduced slowly in phase with the swing, it can be made to swing higher or lower. If the applied force is out of phase, the swing motion is disrupted and becomes erratic. In the Pontzen mechanism, in general, the swing gains energy in a random walk; the swing's energy can never be reduced to zero.} As demonstrated by an idealized simulation \citep{pon12n}, the irreversibility thus arises not from dynamical differences but statistical differences between the blow-out and recollapse events. Any amount of reaccretion or recollapse does not return a disc to its (vertically) thinner state.

Fig.~\ref{f:cusp} investigates whether the density cusp residing in the NFW dark matter halo, already weakened by the dominant central disc, can be further softened or removed by the Pontzen mechanism. The evolution of the density profile (upper plots) is shown for the usual gas fractions. Many of the time curves overlap. The effect is really only detectable at the highest gas fractions, in the sense that the ratio of the density profile, after being divided by the starting profile, shows a clear depression at late times near $R=0$; even then, the suppression is at worst a factor of two. So total destruction of the NFW cusp is not achieved, unlike the case for dwarf galaxies with their much weaker central potentials. This finding is compatible with the absence of cusps at our targeted halo mass ($10^{11}$ M$_\odot$) in modern cosmological simulations of galaxy formation \citep{cha15,tol16}.


\begin{figure}[!htb]
\centering
\includegraphics[width=0.8\textwidth,trim=0 0 0 0.6cm, clip]{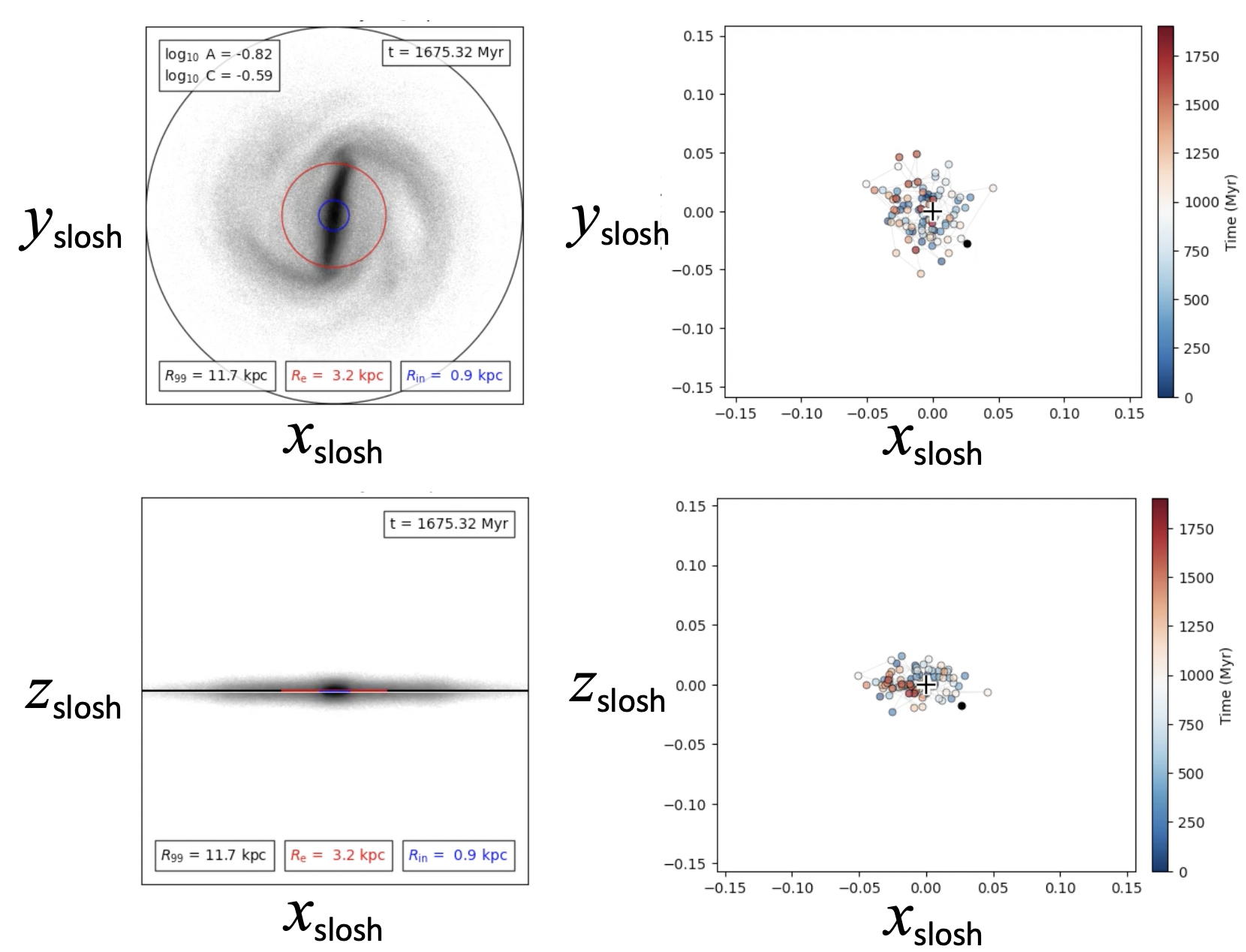}
    \caption{{\tt Nexus} simulation for an $f_{\rm disc}=0.5$ stellar disc with no cold gas component ($f_{\rm gas}=0$\%). The top left and bottom left panels are the $x-y$ and $x-z$ slosh projections respectively at a fixed timestep.  The baryon sloshing is defined in terms of the separation of the minimum of the total gravitational potential ($\Phi_{\rm tot}(0,0)$ indicated by the cross at the origin) and the baryon potential ($\Phi_{\rm bary}(0,0)$). (The boxed information concerns the disc asymmetry and concentration, a topic discussed in our companion paper, \citealt{bla25b}. The physical scales of the 3 circles are indicated in the lower key). The top right and bottom right panels are the $x-y$ and $x-z$ slosh projections respectively, but highly magnified compared to the LHS panels. A small amount of `baryon sloshing' ($\lesssim 50$ pc) is expected due to the discreteness of the disc, as demonstrated. The colour coding indicates the baryon centroid shifts with the passage of time, as indicated. }
   \label{f:slosh0}
\end{figure}
\begin{figure}[!htb]
    \centering
\includegraphics[width=.49\textwidth,trim=0 3cm 0 3cm, clip]
{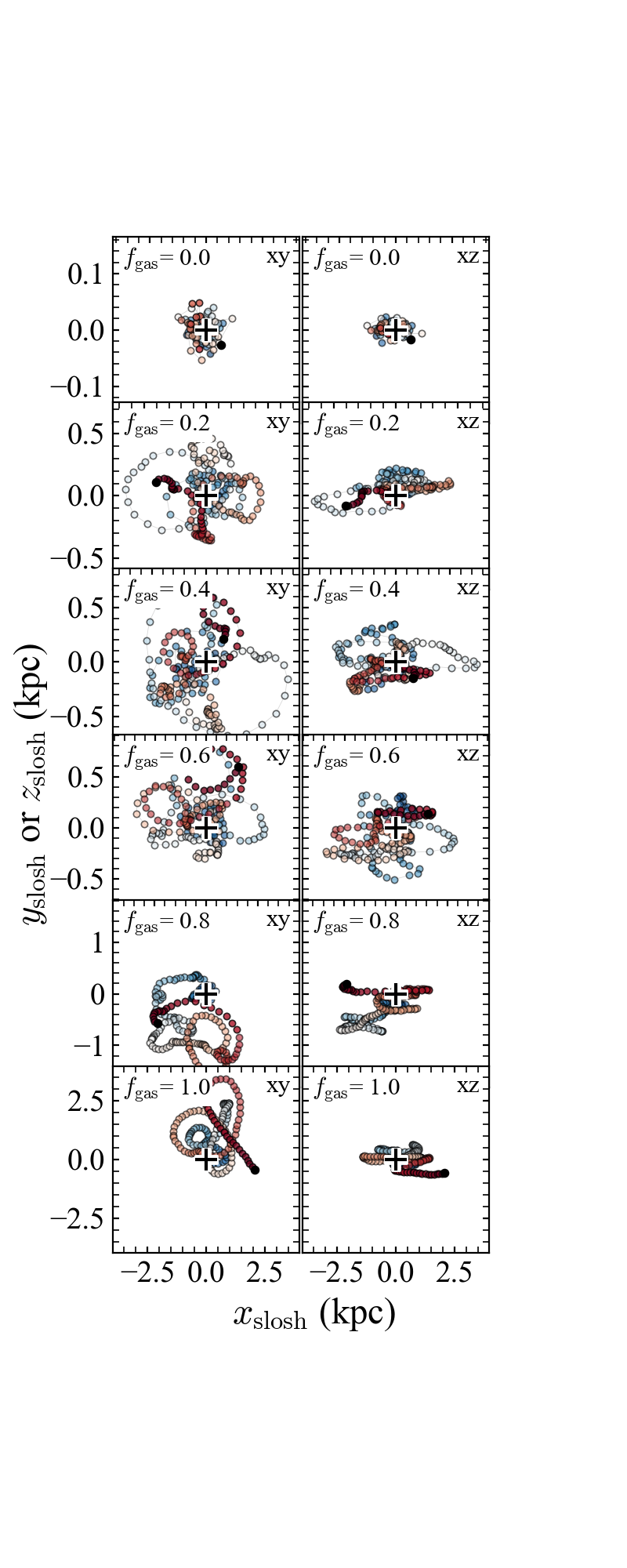}
\includegraphics[width=.49\textwidth,trim=0 3cm 0 3cm, clip]{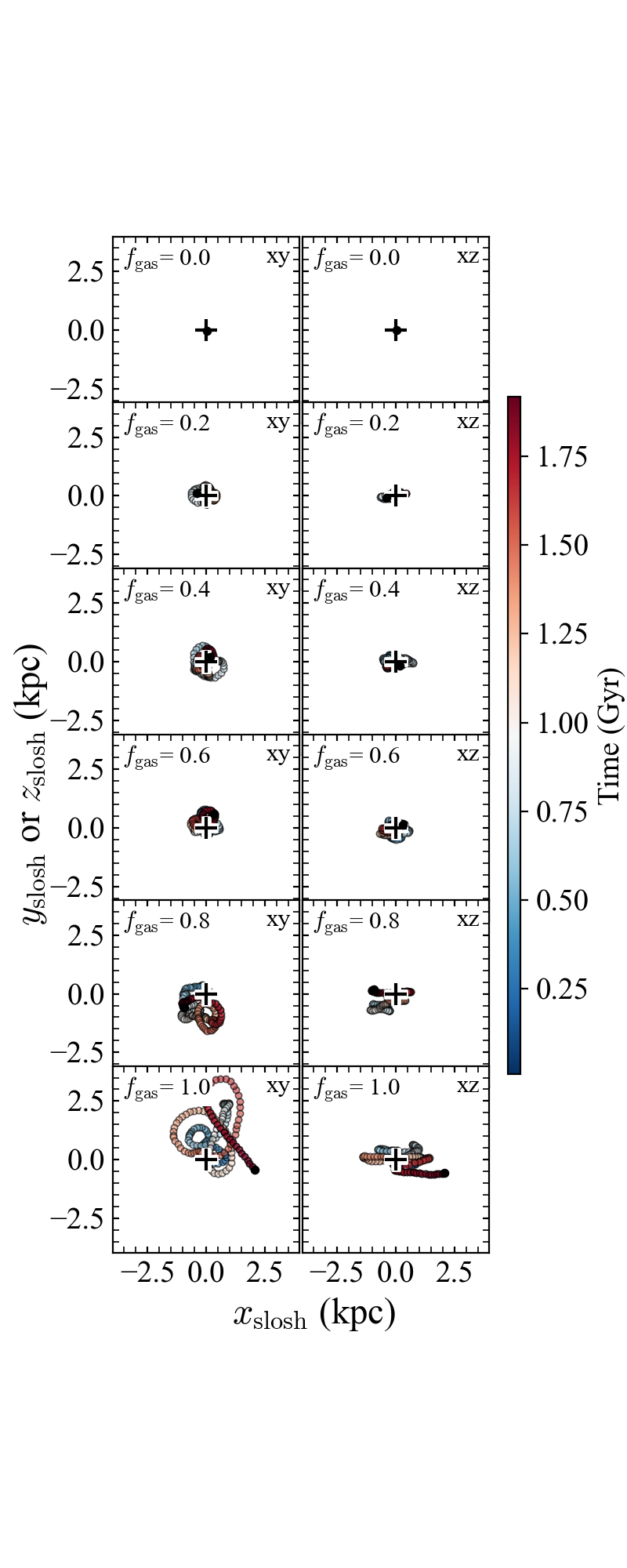}
\caption{{\tt Nexus} simulations in the presence of halo accretion for an $f_{\rm disc}=0.5$ stellar disc for different values of $f_{\rm gas}$, as indicated. {\it Left panel:} $x-y$ (left column) and $x-z$ (right column) projections of the sloshing action. Note that the physical scale of each plot is growing with increasing $f_{\rm gas}$.
{\it Right panel:} $x-y$ (left column) and $x-z$ (right column) projections of the sloshing action.  Note that the physical scale of each plot is fixed for all $f_{\rm gas}$.
The baryon sloshing is defined in terms of the separation of the minimum of the total gravitational potential $\Phi_{\rm tot}(0,0,0)$ (indicated by a cross) and the baryon gravitational potential, $\Phi_{\rm bary}(0,0,0)$. The colour coding indicates the passage of time, as indicated by the RHS bar, extending from $z\approx 3.3$ to $z\approx 1.5$. Internal to each plot, the horizontal and vertical scales are always matched.}
   \label{f:slosh1}
\end{figure} 

\subsection{Sloshing behaviour} \label{s:slosh}
In the {\tt Nexus} simulations, the powerful starburst events drive turbulence across the gas-rich disc through coupling of the mechanical and radiative luminosity arising from star formation. These scattered events, both in time and location, drive global oscillations throughout the disc potential.\footnote{There is a useful analogy here with the Sun. The turbulent motions in its convective zone create random pressure perturbations, which act like `kicks' to the surrounding plasma \citep{hou15} The randomness of the turbulence excites a broad range of discrete modes simultaneously. The acoustic waves generated by convection can reinforce each other through resonance, leading to standing waves that persist. These standing waves correspond to eigenmodes of oscillation with specific frequencies and spatial patterns. Each mode is also damped over time by dissipative processes, but because of the constant churning of convection, new energy is continually injected into the modes, maintaining their amplitudes over long periods. Detectable oscillations induced by these p-modes form the basis of helioseismology and, for stars more generally, asteroseismology \citep{aer21}. Similar processes occur in gas-rich galaxies, but these excite a continuum of (van Kampen) modes rather than standing waves excited by discrete modes.}
The global pressure modes driven by strong impulsive  fluctuations arising in our simulations are not `normal' modes that lead to long-lived standing waves. 
In galaxies, these oscillations would decay in $\lesssim 30$ Myr without the turbulence maintained by ongoing feedback. Interestingly, \citet{WidrowEtAl2012} has used the neologism `galactoseismology' to describe global oscillation modes of the Milky Way\footnote{For a brief history, see \citet{bla21e}.}, but this is a specific reference to the response of the stellar disc. These so-called van Kampen modes \citep{bin25} do not exhibit collective oscillations with a single frequency, as normal modes do, but rather form a continuum of phase-space oscillations. These modes can occasionally reinforce each other (e.g. Gaia phase spiral; \citealt[][]{ant18}) or, alternatively, phase-mix away any coherent behaviour.

\begin{figure}[!htb]
    \centering

\includegraphics[width=1.05\textwidth]{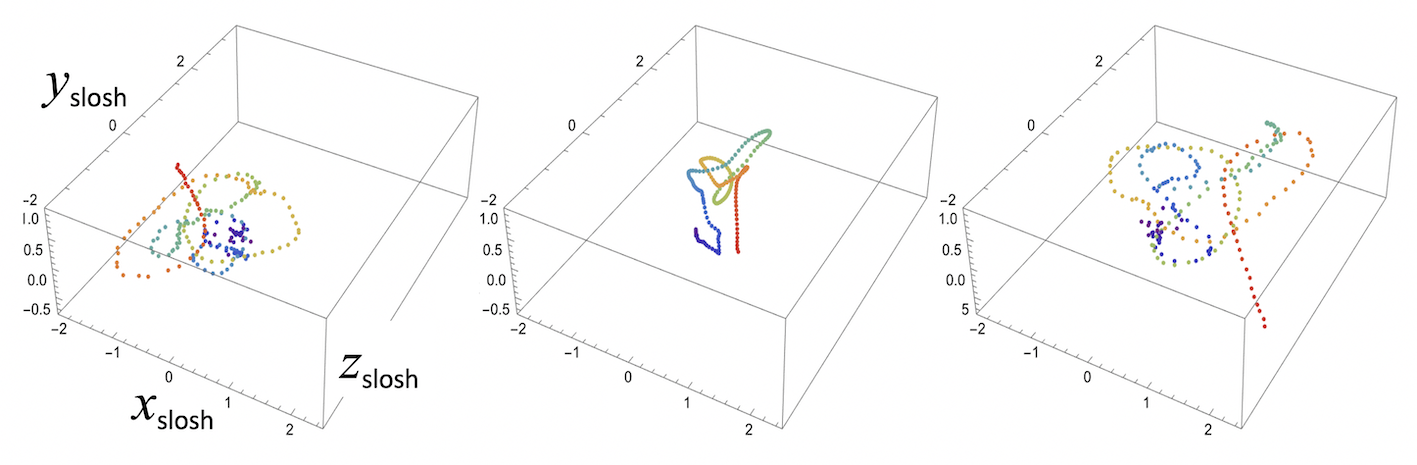}
\caption{The 2D slosh projections in Fig.~\ref{f:slosh1} can be rendered in 3D to reveal the erratic movement. The tracks all refer to the $f_{\rm gas}= 100\%$ halo accretion case: (Left) Motion of $\Phi_{\rm tot}(0,0,0)$ with respect to the inertial reference frame; (Middle) Motion of $\Phi_{\rm bary}(0,0,0)$ with respect to the inertial reference frame; (Right) Motion of $\Phi_{\rm bary}(0,0,0)$ with respect to $\Phi_{\rm tot}(0,0,0)$.
The slosh axes are in units of kpc in a comoving frame. Evidently, for the baryons, there are frequent major reversals over the 2 Gyr simulation. The colour coding indicates the passage of time, from blue to red over this timespan. 
}
   \label{f:slosh3D}
\end{figure} 

Small-scale baryonic oscillations with respect to the dark matter halo have long been recognized \citep[q.v.][]{kuh13,Joshi24}. These can be induced by mergers through momentum recoil, or excessively clumpy, spinning or tumbling dark matter halos. Furthermore, as recognized by \citet{lau21}, an N-body model initialised from the distribution function of an equilibrium model quickly develops macroscopic fluctuations in excess of what is expected from Poisson noise. We confirm this behaviour on 50~pc scales in Fig.~\ref{f:slosh0} for a {\tt Nexus} simulation that incorporates a stellar disc component free of gas ($f_{\rm gas}=0$\%).

The baryon sloshing is defined in terms of the separation of the minimum of the total gravitational potential ($\Phi_{\rm 0,tot}$ indicated by the cross) and the baryon potential ($\Phi_{\rm 0,bary}$) given by equations~\ref{e:tpot} and \ref{e:bpot} respectively.
In Fig.~\ref{f:slosh1}, we repeat the exercise carried out in Fig.~\ref{f:slosh0} to show the sloshing behaviour for $f_{\rm gas}=20,40,60,80,100$\%. The ``post-processed'' movies at our website \href{http://www.physics.usyd.edu.au/turbo\_disks/}{http://www.physics.usyd.edu.au/turbo\_discs/} clearly show the baryon sloshing in the inertial frame. See, for example, the $x-y$ projection of the newly formed stars (left middle panel) in {\tt fd50\_fg100\_ac}; the emerging stellar bar soon drifts away from the origin and follows a complex track with multiple loops.

In Fig.~\ref{f:slosh1}, notice that the vertical migration is half that of the radial migration in all cases, including the $f_{\rm gas}=0\%$ case in Fig.~\ref{f:slosh0}. In Fig.~\ref{f:totalpot}, the massive disc flattens the total gravitational potential along the vertical axis. The vertical restoring force is stronger than the radial force, and thus provides more resistance to the vertical drift of the baryons.
The total (effective) gravitational potential (Fig.~\ref{f:totalpot}) can be approximated with
\begin{equation}
    \Phi_0(R,z) \approx \Phi_1(0,0) + \frac{1}{2}\kappa^2 R^2 + \frac{1}{2}\nu^2 z^2
\end{equation}
where the natural harmonic frequencies of the smooth potential are evaluated at $z=0$ in both instances. For simplicity, we have dropped the $\phi$ ordinate even though the simulated potential is rarely axisymmetric. Thus
\begin{eqnarray}
    \kappa^2(R) &=& \left(\frac{\partial^2\Phi_0}{\partial R^2}\right)_{z=0} \;, \\
    \nu^2(R) &=& \left(\frac{\partial^2\Phi_0}{\partial z^2}\right)_{z=0}.
    \label{e:freq}
\end{eqnarray}
For small perturbations, the restoring force is
\begin{equation}
    {\bm F}_{\rm rest} = M_{\rm halo} \frac{d{\bm V}_{\rm disc}}{dt} \approx -M_{\rm halo} (\kappa^2 R\; \mathbf{i} + \nu^2 z\; \mathbf{k})
\end{equation}
where $M_{\rm halo}$ is the total halo mass (dark matter $+$ hot coronal gas), and ${\bm V}_{\rm disc}$ is the disc's velocity with respect to the total gravitational potential. Thus, as expected, since $\nu > \kappa$, the sloshing perturbations for all simulations is more compressed along the vertical axis when compared with in-plane motions. 

This fact has important implications for the physical consequence of sloshing. For example, the momentum recoil kicks that push the gas around are essentially isotropic to first order, with no preferred direction. While we do see occasional evidence for oscillatory behaviour, the overall effect is more chaotic due to strong stochastic impulses acting in three dimensions. \citet{Joshi24} incorporate an additional damping term to the restoring force, i.e. 
\begin{equation}
    {\bm F}_{\rm damp} = -\beta M_{\rm halo}{\bm V}_{\rm disc}
\end{equation}
to account for the dynamical friction from the unseen dark matter and hot baryonic halo. The factor $\beta \approx 4\pi G^2 \rho_{\rm halo}(R) M_{\rm halo} b /\sigma_{\rm halo}(R)^3$ where the variable $b\approx \frac{1}{4}$ near $R\approx 0$ and goes to zero at infinity, and $\rho_{\rm halo}$ and $\sigma_{\rm halo}$ are the density and velocity dispersion of the halo in the vicinity of the disc.
We arrive at 
\begin{equation}
    {\bm F}_{\rm rest} \approx -M_{\rm halo} (\kappa^2 R\; \mathbf{i} + \nu^2 z\; \mathbf{k}+\beta {\bm V}_{\rm disc})
\end{equation}
This can explain why, in most cases, the disc centroid appears to settle to the $z=0$ plane of the total gravitational potential. Furthermore, in some of the simulations (e.g. $f_{\rm gas}=40$\% in Fig.~\ref{f:slosh1}), the baryonic centre executes a near-circular wide orbit about the total potential, which may arise from the orbit-circularising effect of dynamical friction \citep{jia00}. We return to these issues in the next section.

In Fig.~\ref{f:slosh3D} (Left), the 2D slosh projections shown in Fig.~\ref{f:slosh1} are rendered in 3D to reveal the erratic (Brownian) nature of the movement. Here, we show the halo accretion case for $f_{\rm gas}=100$\%. Evidently, there are frequent major reversals over the 2 Gyr simulation. 
A gallery of radial and kinematic signatures for all $f_{\rm gas}$ is presented in App.~\ref{s:signatures}.

\section{The origins of sloshing} \label{s:pert}

\subsection{Theory vs. simulations}
The dynamics of stellar discs is a remarkably complex process, with only restricted analytic cases that are well understood \citep{bin25,ham25}. So to what extent can we trust N-body simulations to trace the growth of instabilities and non-linear perturbations realistically?  One of the most powerful analytic theories to address this issue is the (local) shearing-sheet model presented by \citet{jul66} for a razor-thin 2D disc. Here, a perturbation in the distribution function (DF) can be followed in both density space and velocity space. A local disturbance generates a leading wake that is ``swing-amplified'' into a trailing wake. These density perturbations grow as a consequence of gravity, shear and rotation \citep{too64a}, but eventually dissolve unless sustained by a perturbing (e.g. external) force. Even when the density enhancement has disappeared, its influence is seen in velocity space long afterwards but eventually fades from view due to phase-mixing.

In the shearing sheet approximation, the swing-amplified modes are analysed in a frozen, co-rotating frame where both the global and local rotation (due to shear and vorticity) are removed. The model treats stars that are on near-circular orbits, although corrections can be made for mildly non-circular orbits \citep{bin20}. In such a frame, all curvature is removed and spiral features can be treated as time-dependent plane waves, typically via a local density (or surface density) $-$ gravitational potential pair of coupled functions.

\subsection{Jolting the disc $-$ the importance of self-gravity}

So how does the shearing sheet respond to a powerful jolt? This line of enquiry is revealing because it shows that an impulse experienced by a stellar disc is amplified by its self-gravity for all $Q_\star$ $(=\kappa\sigma_R/3.36 G\Sigma_0)$ values relevant to {\tt Nexus} galaxies. Here, $\Sigma_0$ is the local undisturbed surface density, $\kappa$ is the epicyclic frequency and $\sigma_R$ is the {\it radial} velocity dispersion. $Q_\star$ is a dimensionless quantity that measures the local axisymmetric stability in the stellar disc against collapse or fragmentation; non-axisymmetric growing modes are still allowed. The amplifying effect is strongest at intermediate wavelengths (see below), becoming weaker at long wavelengths, while short wavelengths are suppressed below a critical wavelength $\lambda_{\rm crit} = 2\pi/k_{\rm crit}$, for which $k_{\rm crit}=\kappa^2/2\pi G \Sigma_0$. The characteristic epicyclic radius is roughly $\sigma_R/\kappa$.  These physical dimensions are on kiloparsec scales, e.g. $k_{\rm crit}\approx (2\pm 1\rm\; kpc)^{-1}$, such that for example $\lambda_{\rm crit} \sim 10$ kpc near the Solar circle today.

In the shearing-sheet approximation \citep{jul66}, the coordinate system is centred on a spiralling wave's corotation radius. In Galactic cylindrical coordinates ($R,\phi,z$), individual stars have velocities ${\bm V}$ $=$ ($V_R$, $V_\phi$, $V_z$) and oscillation frequencies ${\bm \Omega}$ $=$ ($\Omega_R,\Omega_\phi,\Omega_z$) $=$ ($\kappa,\omega,\nu$). Following \citet{bin20}, we introduce a cartesian wave vector ${\bm k}$ that corotates around a point defined by the radius vector $\bm R$ such that 
\begin{equation}
    \vert {\bm k} \vert = \vert k_x \bm i + k_y \bm j\vert = k(t) = k_y\sqrt{1+(2{\cal A}(t-t_0))^2} \, ,
    \end{equation}
where $k_y$ is the tangential wavenumber, and ${\cal A} = -\frac{1}{2}\frac{d\Omega_\phi}{d\ln r}$ is Oort's constant that quantifies the shear rate. The wave vector ensures that the perturbation's wave crests track from leading ($t<t_0, k_x < 0$) to radial ($t=t_0, k_x=0$) to trailing waves ($t>t_0, k_x >0$), where the amplitude peaks near $t=t_0$.
A suitable potential $-$ surface density pair is
\begin{eqnarray}
    \Sigma(\bm x, t)&=&\Sigma_0 + \Sigma_1 \cos(\bm k.\bm x) \\
    \Phi_1(\bm x, z, t) &=& -2\pi G \frac{\Sigma_1}{k(t)} \cos(\bm k.\bm x)\; e^{-k(t)/\vert z \vert} \, ,
\end{eqnarray}
where $\bm x$ describes the $x-y$ plane, and $\Sigma_0$ and $\Sigma_1$ are the undisturbed and wave-disturbed density fields. 

\begin{figure}[!htb]
\centering
\includegraphics[width=0.8\textwidth,height=0.5\textwidth]{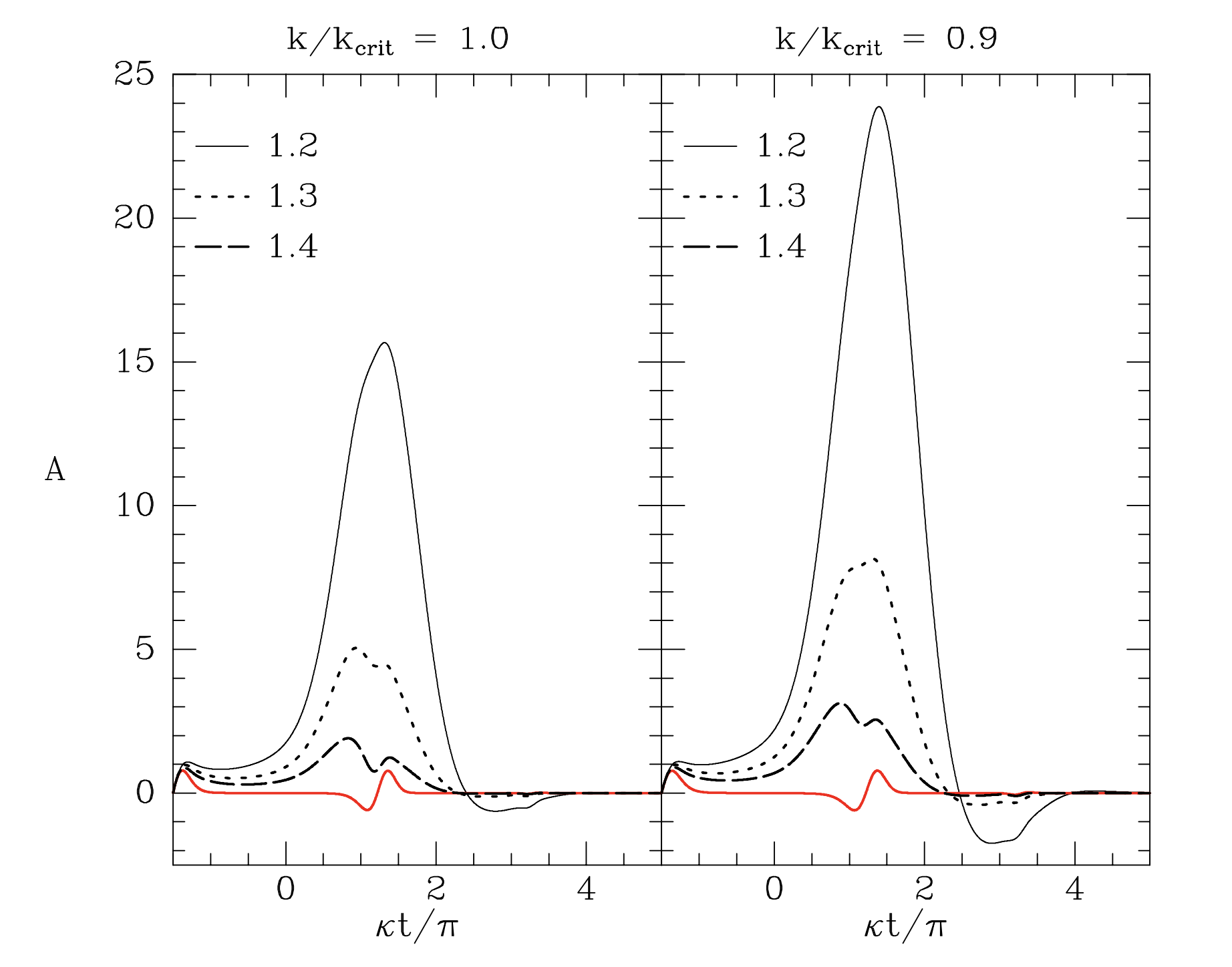}
    \caption{Evolution of wave amplitudes $A(t)=\tilde{\Sigma}_1(t)$ in the presence of an impulse, obtained by solving the iterative integral in Eq.~\ref{e:jt}, where the tilde indicates a surface density amplitude normalized to evolution without gravity. The horizontal axis shows time evolution in units of $\kappa/\pi$, consistent with shearing-sheet analysis \citep{jul66}, starting at $\kappa t_i=-\frac{3\pi}{2}$.
    (Left) The response of a Mestel disc to a sharp impulse for $Q=1.2,1.3,1.4$, demonstrating the increasing importance of self-gravity to the disc's response \citep{bin25}. The red curves illustrate the weak disc response when self-gravity is neglected.
    (Right) This repeats the left panel but for a longer-wavelength disturbance; only wavelengths shorter than $k=k_{\rm crit}$ are stabilised by the initial $Q$ value. The range of $Q$ values shown describes the {\tt Nexus} discs for most of their evolution.
    }
   \label{f:mestel}
\end{figure}

In Fig.~\ref{f:mestel}, we show the response of a razor-thin 2D disc to a strong local impulse for different values of the $Q$ parameter, from $Q=1$ (marginally stable, high self-gravity) to $Q=1.4$ (most stable, low self-gravity).
We initiated the impulse at a time $t=t_i$ {\it before} $t=t_0$; the perturbed distribution function is $f_1(t_i)=0$ at the start. The impulse is such that the surface density from the perturbation is $\tilde{\Sigma}_e(t,t_i) = \delta(t-t_i)/\kappa$. Time is needed for the perturbation to build, but if that time is set too far in the past, the disturbance can phase-mix away before it has a chance to build.
From the so-called JT equation, the renormalized surface-density evolution is given by
\begin{eqnarray}
    \tilde{\Sigma}_1(t) &=& \int_{t_i}^t {\cal K}(t,t^\prime) (\tilde{\Sigma}_1(t^\prime) + \tilde{\Sigma}_e(t^\prime))\; dt^\prime \\
    &=& {\cal K}(t,t_i) + \int_{t_i}^t {\cal K}(t,t^\prime) \tilde{\Sigma}_1(t^\prime)\; dt^\prime \, .
    \label{e:jt}
\end{eqnarray}
The JT kernel is
\begin{equation}
    {\cal K}(t,t_i) =\frac{ 4{\cal F}(t,t_i)}{k(t)\exp(0.57Q^2b^2)} \, ,
\end{equation}
for which $b$ depends on $k_y/k_{\rm crit}$, and ${\cal F}(t,t_i)$ depends on geometric factors involving ${\cal A}$, $\kappa$ and $\Omega$ \citep{bin20}. The first RHS term in Eq.~\ref{e:jt} arises from the impulse and the second integral term is the temporal, accumulated response to the disc's self-gravity. As observed in the simulations, the strong impulses excite a continuum of long-wavelength waves that carry energy and momentum throughout the stellar disc, the gas disc and the dark-matter halo.

In the {\tt Nexus} simulations, away from the galaxy's centre, $Q$ is in the range $1.2-1.5$, leading to swing-amplification factors $A$ as high as $10-60$ compared to discs without self-gravity \citep[cf.][]{aumer16}. \citet{bin20} showed that the biggest amplification occurs around $\lambda/\lambda_{\rm crit}\approx 2\pm 1$ \citep[see also][]{too81}. Thus, it is entirely reasonable that a succession of disc jolts leads to a strong reaction from the large-scale stellar disc (e.g. spiral arm excitation, internal disc heating) on tens of kiloparsec scales \citep[q.v.][]{ham25}. Once the momentum-injecting starbursts die down, the disc ceases to slosh, as observed in the simulations.

\begin{figure}[!htb]
\centering
\includegraphics[width=0.8\textwidth]{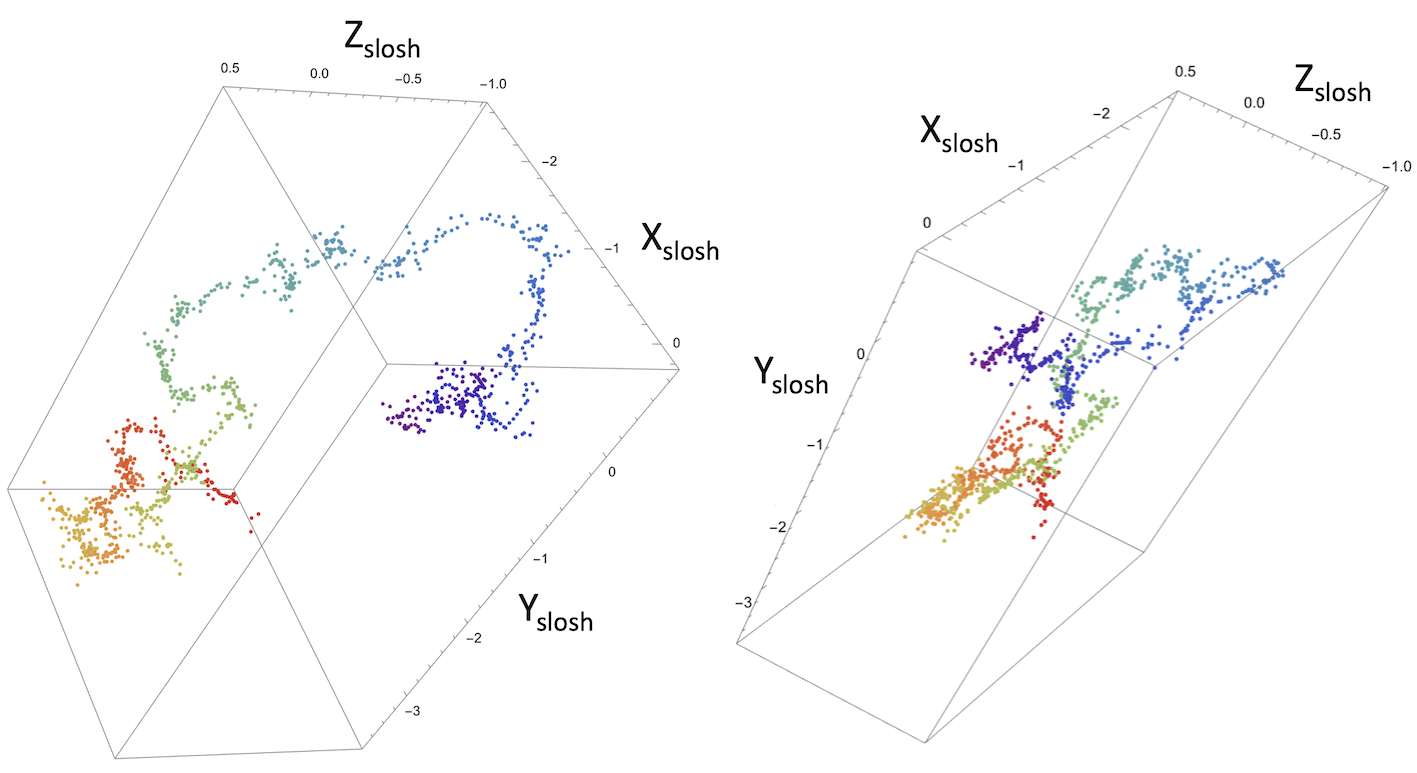}
    \caption{Two different perspectives of the same Brownian motion track in a moderately flattened, spherical gravitational potential. The slosh coordinates (in kpc) are equivalent to those presented in Fig.~\ref{f:slosh1}. The colour scheme once again shows the evolution over 2 Gyr (blue $\rightarrow$ red). The track is similar in character to the baryon sloshing tracks in Figs.~\ref{f:slosh1} and \ref{f:slosh3D}.
    }
   \label{f:brownian}
\end{figure}
\subsection{Brownian motion}
\label{s:brown}

We separate the phenomenon and consequences of baryon sloshing in the total gravitational potential into two parts, which can both be explored with the Fokker-Planck equation
using a probability density function $P(\bm R -\bm R_0, t)$, for which $\bm R$ is the radius vector from an initial position $\bm R_0$. First, the baryon potential minimum moves around stochastically with what is effectively damped Brownian motion.
We demonstrate this property with a simple model. 
Secondly, the internal response of the baryons is subject to energy loss and energy gain in different measures due to the changing underlying potential. In 
a later paper, we treat the internal disc evolution (heating and cooling) separately as a diffusion problem in action space.

For the simple model,
there are ``coasting'' phases where the simulated galaxy can move on a trajectory, but with associated noise due to the stochastic nature of the starburst activity. The overall damping is due to the simulated galaxy moving in a deep potential well and dynamical friction due the disc's motion in a live dark matter halo. On occasion, the entire galaxy can bob up and down with damped quasi-periodic motion before reverting to an erratic trajectory. Here we stress the Brownian motion behaviour because it explains why the disc heating is essentially an isotropic process (Sec.~\ref{s:internal}).

To simulate the track of a particle, Langevin's equation is used to describe the motion subject to (over-damped) Brownian motion, i.e. $\gamma\bm \dot{\bm R} \approx \bm F(\bm R)+\bm \eta(t)$. Here, $\bm F$ is the continuously applied force (e.g. galactic potential gradient), $\eta$ supplies a random force term and $\gamma$ is a damping constant.
In our N-body simulations, the damping is mostly due to dynamical friction, i.e. energy loss to the live dark matter halo. To add an extra layer of realism, we computed the baryon motion in a moderately flattened, spherical gravitational potential (${\bm F} = -\nabla\Phi_0$) that mimics the potential in Fig.~\ref{f:totalpot}. This weakly biases each step $d{\bm R}$ along the radius vector towards the centre, and compresses the motion in $z$. An example of such a track is shown in Fig.~\ref{f:brownian} from two perspectives; most realizations capture the essence of Figs.~\ref{f:slosh1} and \ref{f:slosh3D}.
In the absence of external forces, we can mimic isotropic diffusion in three dimensions using a 3D Gaussian random variate such that ${\bm \eta}(t) = {\cal G}[\bm R(t)-\bm R_0,{\bm \sigma}]$ where $\vert{\bm \sigma}\vert \propto \sqrt{D_2 t}$. 
For an ensemble, we find:
(i) the first moment is $\langle \eta_i(t)\rangle = 0$; (ii) the second moment is
$\langle \eta_i(t)\eta_j(t^\prime)\rangle =  2{\cal D}_2\delta_{ij} \delta(t-t^\prime)$ where $\delta_{ij}$ is the Kronecker delta and $\delta$ is the Dirac delta function. The latter ensures the covariance of the (Wiener) noise process
is uncorrelated. ${\cal D}_2$ is a diffusion coefficient that we chose arbitrarily, along with the timesteps, to match the amplitude of motion in kiloparsecs seen in the N-body simulations.


\begin{figure}[!htb]
    \centering
\includegraphics[width=.35\textwidth]{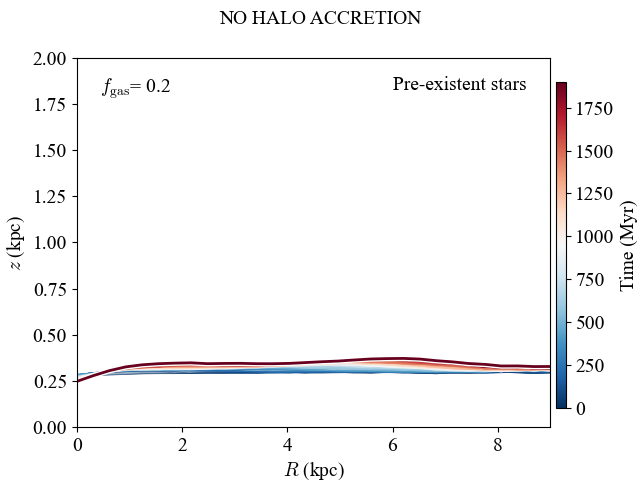}
\hspace{0.5cm}
\includegraphics[width=.35\textwidth]{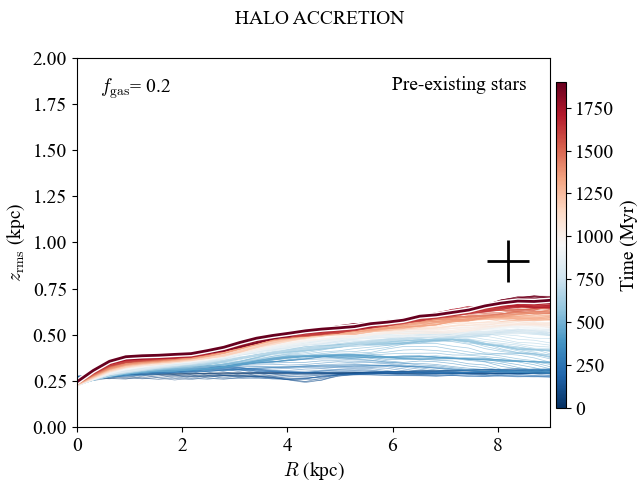}
\includegraphics[width=.35\textwidth,trim=0 0 0 1.1cm, clip]{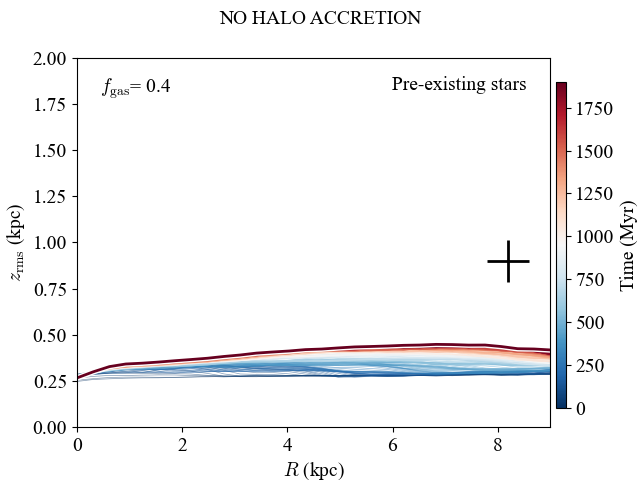}
\hspace{0.5cm}
\includegraphics[width=.35\textwidth,trim=0 0 0 1.1cm, clip]{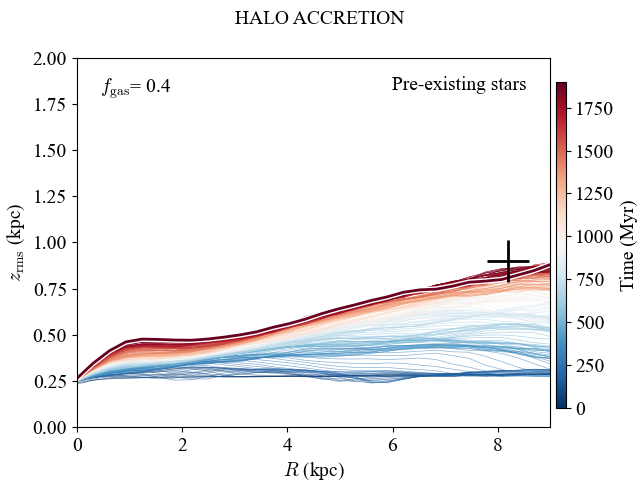}
\includegraphics[width=.35\textwidth,trim=0 0 0 1.1cm, clip]{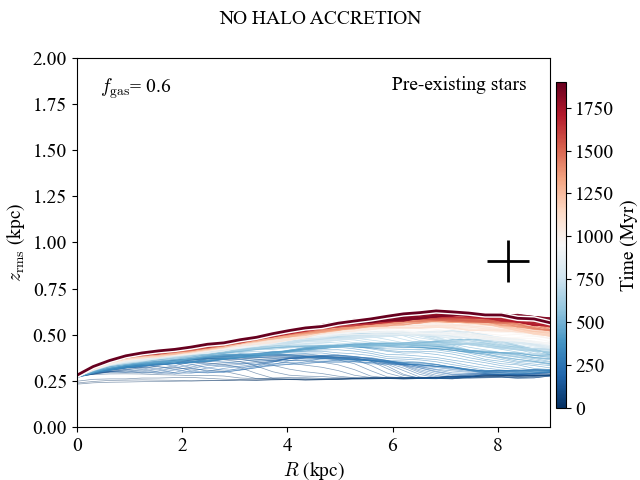}
\hspace{0.5cm}
\includegraphics[width=.35\textwidth,trim=0 0 0 1.1cm, clip]
{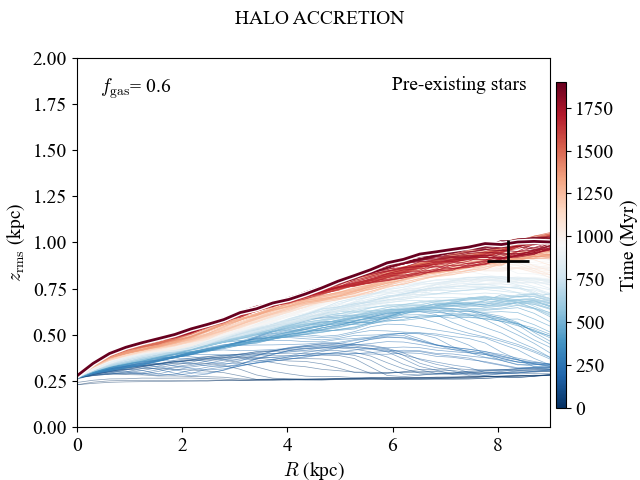}
\includegraphics[width=.35\textwidth,trim=0 0 0 1.1cm, clip]{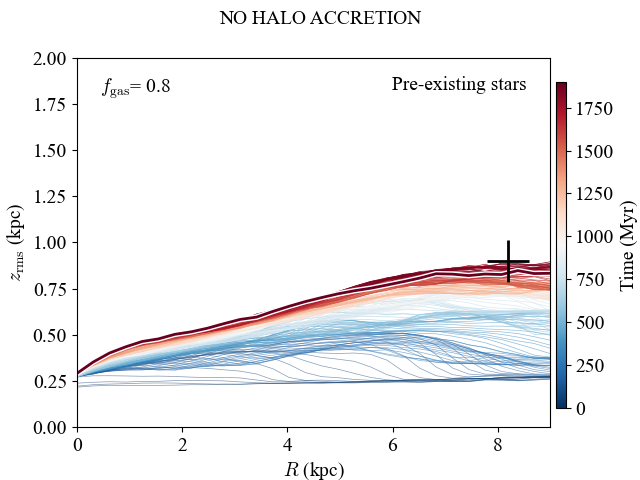}
\hspace{0.5cm}
\includegraphics[width=.35\textwidth,trim=0 0 0 1.1cm, clip]
{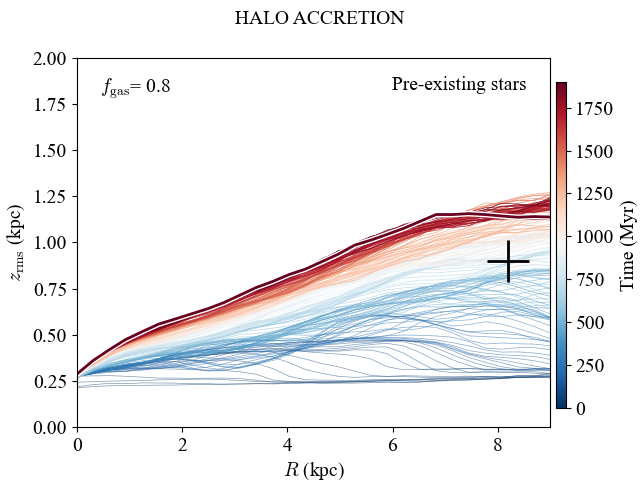}
\includegraphics[width=.35\textwidth,trim=0 0 0 1.1cm, clip]{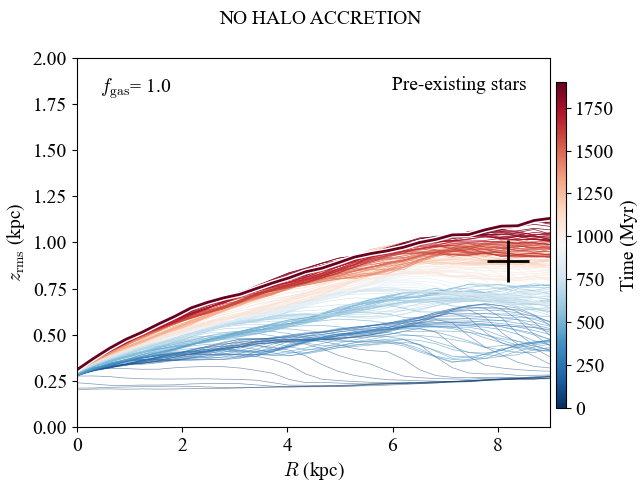}
\hspace{0.5cm}
\includegraphics[width=.35\textwidth,trim=0 0 0 1.1cm, clip]
{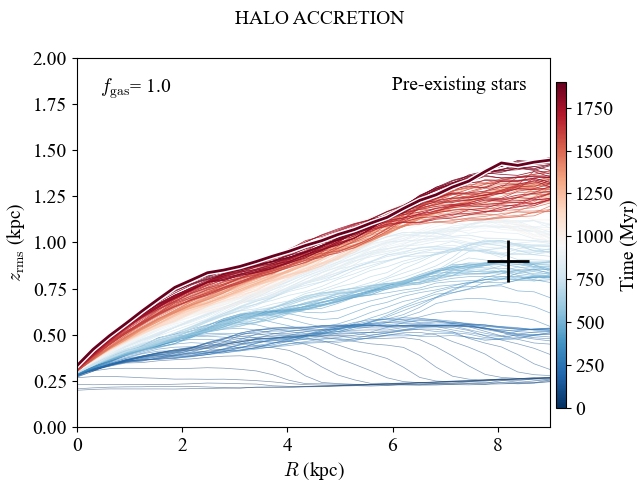}
\caption{The rms $z$-scale height for pre-existing (old) stars as a function of radius $R$ (in kpc) for each value of $f_{\rm gas}$. The left panel is for no-halo accretion cases; the right panel is for halo accretion models. In all plots, the colour coding indicates the time evolution, as indicated. The scale height of the $\alpha$-rich thick disc today at $R_0=8.2$ kpc is indicated by the cross \citep{bla16}. Note that the vertical scale has a fourfold stretch compared to the horizontal axis. Gas-rich models are more able to reproduce the $\alpha$-rich disc within $1.5-2$ Gyr.}
   \label{f:thicken2}
\end{figure}

\begin{figure}[!htb]
    \centering
\includegraphics[width=0.45\textwidth]{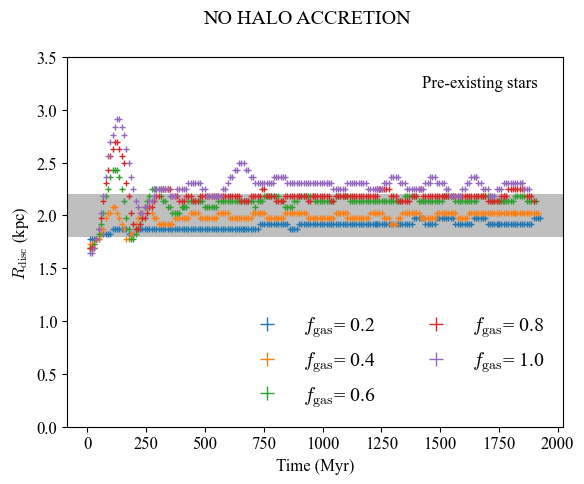}
\includegraphics[width=0.45\textwidth]{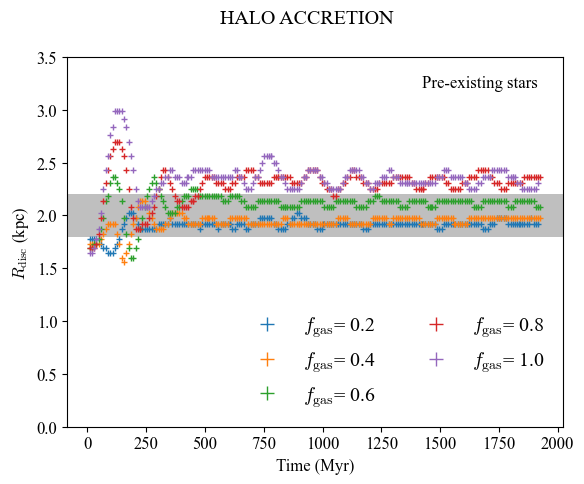}
\includegraphics[width=0.45\textwidth]{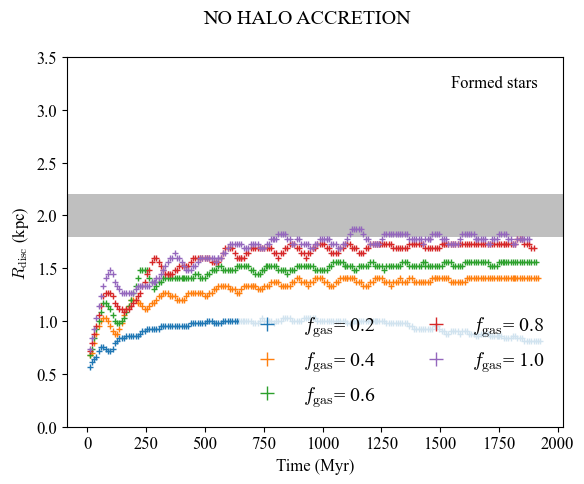}
\includegraphics[width=0.45\textwidth]{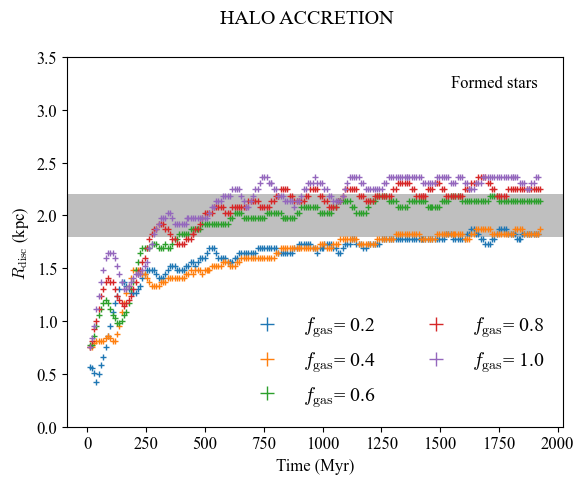}
\caption{(Top) The radial scale length for pre-existing stars as a function of time (in Myr) for each value of $f_{\rm gas}$ indicated in the key. The left panel is for no-halo accretion models; the right panel is for halo accretion models. (Right) The same quantity but for newly formed stars as a function of time (in Myr) for each value of $f_{\rm gas}$ indicated in the key.
In all panels, the grey band indicates the radial scale length of the thick disc today, i.e. $R^T_{\rm disc}=2.0\pm0.2$ kpc \citep{bla16}.
}
   \label{f:scalelength}
\end{figure} 

\begin{figure}[!htb]
    \centering
\includegraphics[width=0.9\textwidth, trim=0 0 0 0, clip]{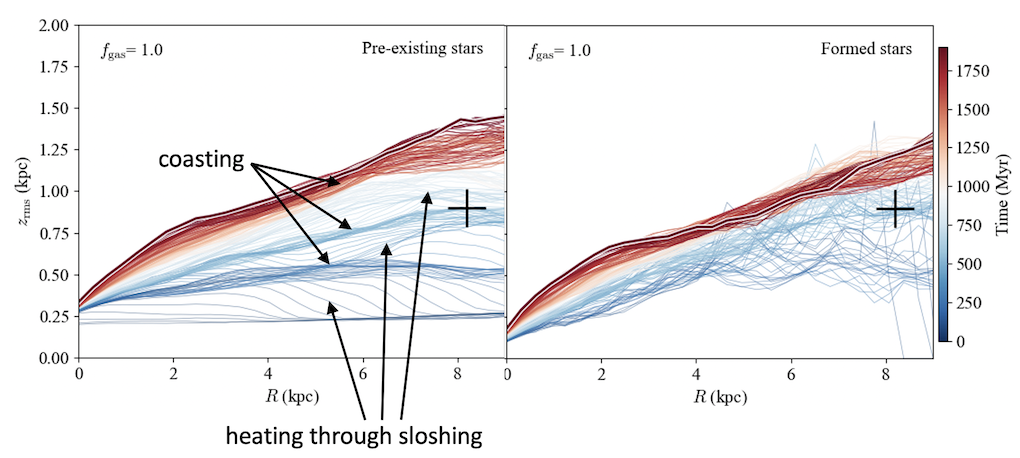}
\caption{
(Left) The $f_{\rm gas}=100\%$ accreting halo case is repeated from Fig.~\ref{f:thicken2} highlights the different heating phases due to rapid reversals along the sloshing track. In practice, we allow for 1\% of the disc mass to be an existing stellar population so that we can trace the impact of the heating.} The $z-$height is starting to saturate after 1.5 Gyr. (Right) This is a repeat of the LHS plot but for stars forming in the gas rather than pre-existing stars. The heating rate is essentially the same. The bottom 10 tracks (100 Myr apart) are removed because of large intrinsic scatter; this is a consequence of small number statistics and early populations being born out of dynamical equilibrium.
Note for both plots that the vertical scale has a fourfold stretch compared to the horizontal axis. The cross indicates the vertical scale height of the Milky Way's $\alpha-$rich disk at the Solar Circle today.
   \label{f:thicken1}
\end{figure} 


\section{Discussion: The consequences of sloshing}
\label{s:disc}

Here we discuss the main consequences of baryon sloshing, at least as we understand them today. The {\tt Nexus} simulations are only run for a few billion years at present, long enough to explore the formation and evolution of the $\alpha$-rich disc. In a later paper, we will trace the evolution of baryon sloshing over the lifetime of the Milky Way. Here, we look at how the vertical and radial scale lengths evolve, and discuss in the context of the sloshing action. We also look at the predicted metallicity trends throughout the $\alpha$-rich disc.


\begin{figure}[!htb]
    \centering
\includegraphics[width=\textwidth]{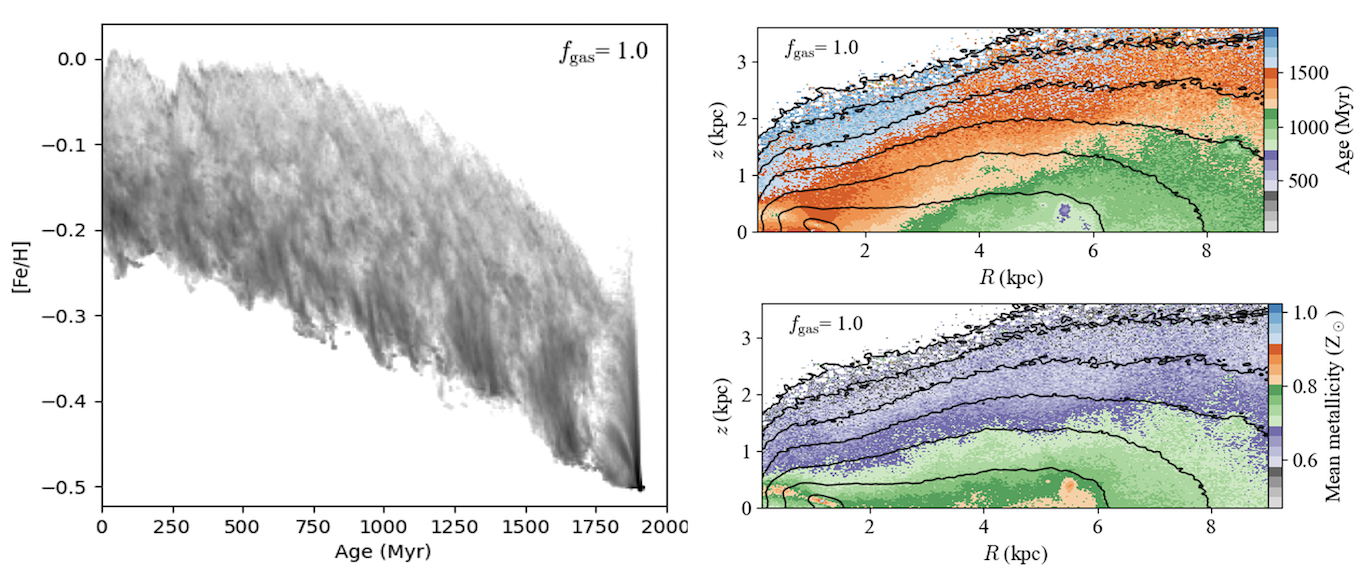}
\caption{Predicted trends at 2 Gyr ($z\approx 1.5$) after disc onset for the halo accreting $f_{\rm gas}=100\%$ model, representative of the high $f_{\rm gas}$ models in general.
(Left) Age-metallicity trend in the simulated $\alpha$-rich disc in the simulated $R-z$ plane, i.e. within a vertical slice that is 100 pc thick along the line of sight.
(Top right) Age distribution in the same plane showing a weak trend in radius but a stronger vertical gradient, such that the youngest stars lie near the plane.
(Bottom right) Metallicity distribution in the same plane showing a weak trend in radius but a stronger vertical stratification, such that the most metal-rich stars lie near the plane. For the RHS figures, representative stellar density contours within the vertical slice are overlaid.}
   \label{f:metals}
\end{figure} 
\subsection{Disc thickening}
\label{s:thickdisc1}

The formation of early thick stellar discs has been a challenge for theory and simulations alike. One class of models invoke stars forming in a turbulent, kinematically hot ISM to generate a thick stellar disc, which can be further heated by mergers or interactions with satellites
\citep{2013MNRAS.436..625S,2017MNRAS.467.2430M,2021MNRAS.503.1815B,van22}. However, this effect is not seen in our higher-resolution {\tt Nexus} simulations. In Sec.~\ref{s:thickdisc2}, we present evidence {\it against} disc thickening arising from {\it in situ} star formation in a turbulent gas disc. 
Although star formation and turbulence are still involved, our mechanism is very different. 

In Fig.~\ref{f:thicken2}, the rate of disc heating increases dramatically with $f_{\rm gas}$. This is shown for pre-existing stars as these are a good tracer of the internal heating. (We present the matching plot for newly formed stars in Sec.~\ref{s:thickdisc2}.) The internal heating continues until the gas fraction drops and the star formation rate follows suit, at which point the internal heating becomes saturated. The high $f_{\rm gas}$ models achieve the $\alpha$-rich thick disc values today. The $f_{\rm gas}=80\%$ and 100\% models overshoot the measured values today, but this is likely to have taken place in the early Milky Way. Once the younger, massive thin disc forms, the pre-existing $\alpha$-rich thick disc must have been compressed by the increased density of the $\alpha$-poor disc.

Another consequence of disc heating is the rise of the vertical stellar dispersion $\sigma_z$ in both pre-existing and newly formed stars with cosmic time. The relevant figures are presented in our earlier paper \citep[see Figs. 6 and 20 in][]{bla24}. In Sec.~\ref{s:internal}, we present a simple model for how this heating arises from the fluctuating gravitational potential.
Evidence for quasi-isotropic heating is the roughly preserved ratios $\sigma_z/\sigma_R$ and $\sigma_\phi/\sigma_R$ across all radii for the pre-existing stars throughout the simulation. Initially, the situation is different for the newly formed stars because their radial distribution exhibits inside-outside growth (Fig.~\ref{f:scalelength}), as observed in our movies. But the heating rate of the younger stars matches that of the pre-existing stars once the formed stars extend across the pre-existing disc
(see Sec.~\ref{s:internal}).

In Fig.~\ref{f:scalelength}, the evolution of the radial scale length is presented over the full simulation for the pre-existing (top) and newly formed stars (bottom).
To combat noisy data, we determine the disc scale length $R_{\rm disc}$ from the radius enclosing half the disc's mass; we divide the latter by 1.67, which is appropriate for a 3D exponential disc. For the pre-existing stars, we see the disc grow radially by no more than about 20\% of its starting dimension. For newly formed stars, the growth is more dramatic. All halo-accretion models achieve the radial scale length observed for the thick disc today \citep{mac17,bla16}, indicated by the grey band. For the non-accreting models, a very high starting $f_{\rm gas}$ value is required.
In contrast, the vertical disc scale height grows by a factor of $1.5-8$ depending on the model (Fig.~\ref{f:thicken2} and App.~\ref{s:thickdisc2}
). This statement is not inconsistent with the claim of isotropic heating: the young disc is initially very cold in the vertical direction ($\sigma_z \sim 5$ km s$^{-1}$) compared to the radial direction ($\sigma_R \sim 60$ km s$^{-1}$).

In Fig.~\ref{f:thicken1}, we highlight a causal connection between the baryon sloshing motion and the progressive vertical heating of the disc. 
The vertical growth goes through different phases of slow and fast growth. 
With reference to the 3D sloshing track for this specific model (Fig.~\ref{f:slosh3D}), we should not be surprised that the overall heating is roughly isotropic when averaged over time. In what follows (Sec.~\ref{s:internal}), a dynamical argument is given in support of this statement.

\subsection{Chemical signatures}
\label{s:chem}

We provide preliminary results on the predicted stellar chemistry for the early thickened disc \citep[see][for details]{tep24,agertz2021}. Some of our ``results'' are not particularly compelling because, like all simulators, we set up the initial conditions for success. Specifically, the initial star formation rates were very high and this guaranteed the production of high-mass stars and supernovae under starburst conditions. These, in turn, produced $\alpha$-enhanced yields ensuring the first generations of stars are $\alpha$-enriched. This signature remains as some of the best supporting evidence that early discs were gas-rich \citep{fre02d,bla16b}. Simulators often present their ``predicted'' old $\alpha$-rich stars as a success of the model, but the first generations tend to emerge as a direct consequence of the initial set-up conditions. Arguably, it is the subsequent generations that are more useful.

In particular, the left panel in Fig.~\ref{f:metals} shows a clear trend in [Fe/H] as time passes. 
This is significant because the thick disc has been described as a snap-frozen, mono-age stellar population (see Sec.~\ref{s:intro} and \citealt{fre02d}), which is inconsistent with our model. Well defined age-metallicity trends have been discussed extensively in the context of the $\alpha$-rich disc \citep{hayw13,hay14,xia22,Mackereth2018,sha21}.
Most recently, this is precisely what is seen for the $\alpha$-rich population in the GALAH/Gaia \citep{sahl22} and APOGEE/Gaia surveys \citep{cerqui25}, both in terms of timescale and metallicity range.

The right panel in Fig.~\ref{f:metals} show $R-z$ cross-sections of the thickened stellar disc at 2 Gyr ($z\approx 1.5$) after the disc formed at $z\approx 3.3$. The upper panel exhibits a clear vertical gradient in stellar age, with the oldest stars furthest from the plane.
The lower panel shows a clear vertical stratification in stellar metallicity with the most metal-poor stars furthest from the plane. These general trends (albeit described by different functions of $R$ and $z$), including the flaring at large radius due to the declining disc potential, are well established for both the $\alpha$-poor and $\alpha$-rich discs \citep{bov12,mac17}.

In both panels, the radial trends are very weak, unlike what is observed in the mass-dominant $\alpha$-poor disc. Radial gradients have been searched for repeatedly in APOGEE data, for example, but never established \citep{hay14,and14}. Currently, we are unable to say whether the lack of a radial gradient reflects the thick disc's formation epoch, or its subsequent evolution. For example, was there an early radial gradient that was washed away by subsequent bar formation? The lack of radial structure makes it difficult to determine a likely birth radius for individual stars using chemical tagging arguments. This question requires more investigation.

We would argue that, on the evidence provided, the correct vertical age-metallicity trends and the non-existent radial trends are successes of the baryon-sloshing model. More extensive analysis is required to establish whether the baryon-sloshing model has unique chemodynamic signatures that we can search for.

\subsection{Internal heating of the stellar disc}
\label{s:internal}
Contemporary discussions of Galactic heating tend to focus on either vertical or in-plane processes, or heat energy lost to the dark matter halo, although 3D disc heating has been discussed \citep{ida93}. This is because the discussion centres on an observed phenomenon and a specific process, e.g. radial abundance gradients and in-plane migration. But here we stress that baryon sloshing must lead to isotropic heating, at least to first order, such that all Galactic coordinates are affected simultaneously. Figs.~\ref{f:thicken2} and \ref{f:scalelength} are evidence enough that vertical and in-plane heating occur concurrently in the simulations. In a later paper, we return to a detailed discussion of 3D heating using a Fokker-Planck diffusion analysis in action space. 

Instead, we use a simple model for how vertical heating of the stellar disc arises from sloshing in a time-dependent background potential. The connection between fluctuating gravitational fields and internal dissipation or heating is a classical problem in statistical physics \citep{gre54,kub57}, and has also been examined in astrophysical dynamics \citep[e.g.][]{nel99,wei01}. To build intuition, consider a galactic disc oscillating vertically within an external potential. If the fluctuation $\delta\Phi$ has a zero mean around the equilibrium position ($z=0$), the equation of motion for a stellar mass $m$ is $m\dot{v}_z=F(z,t)=-\nabla\delta\Phi$. For simplicity, we set $m=1$ since the mass cancels out in the dimensionless form of the analysis. For a stochastic process, the diffusion coefficient is defined as 
\begin{eqnarray}
    {\cal D}&=&
     \frac{1}{2}\frac{d}{dt}\langle(v_z(t)-v_z(0))^2\rangle \\
    &=&\frac{d}{dt}\int_0^t (t-\tau)\:{\cal C}(\tau)\:d\tau
\end{eqnarray}
where the force autocorrelation function is 
${\cal C}(\tau)=\langle F(0)F(\tau)\rangle$. As $t$ goes to infinity, the diffusion coefficient simplifies to
\begin{equation}
    {\cal D}=\int_0^\infty {\cal C}(\tau)\:d\tau \approx \langle F^2\rangle \tau_{\rm corr}
\end{equation}
assuming ${\cal C}$ decays over a correlation time $\tau_{\rm corr}$ and has a variance
$\langle F^2\rangle$. The baryon sloshing decays over timescales of 400-700 Myr for the different models as the gas is converted into stars. We approximate 
$\langle F^2\rangle=(\delta\Phi_z/Z)^2$ where $Z$ is the vertical distance over which the potential is varying.
For $\Phi_z\approx \sigma_z^2$ and $\tau_{\rm dyn}\approx Z/\sigma_z$, assuming the simulation time $T_o$ (2 Gyr) is much longer than $\tau_{\rm dyn}$ ($\approx600/2\pi$ Myr), 
we can approximate the change in the stellar velocity dispersion to be 
\begin{equation}
\frac{\delta\sigma_z^2}{\sigma_z^2} \approx \frac{2}{3} \left(\frac{\delta\Phi_z}{\Phi_z}\right)^2 \frac{\tau_{\rm corr}\:T_o}{\tau^2_{\rm dyn}} .
\end{equation}
as a function of the change in the underlying gravitational potential.

The clearest demonstration of how stellar populations heat with time is presented in \citet[][Fig. 6]{bla24}, drawn from the same data used in Fig.~\ref{f:thicken2}. For the highest gas-fraction models, the pre-existing stars increase their dispersions by factors of $1.5-2$, requiring gravitational fluctuations of $10-20\%$. The young stars, at least those born in a dynamically stable disc, have a colder intrinsic dispersion ($\sigma_z\approx 6-8$ km s$^{-1}$): these increase their dispersions by factors of $3-4$ requiring fluctuations of 50\%. This is an order of magnitude too high to be explained by the underlying dark matter potential or by stars being born out of equilibrium. Instead, we believe the extra heating arises from the large $\delta\Phi_z$ fluctuations due to the unbinding of the disc. We refer the reader to Fig.~\ref{f:pots0} and the discussion in Sec.~\ref{s:unbind} where $30-50\%$ fluctuations are seen in the most extreme cases.  How the unbinding works with the sloshing action is unclear and deserves further study. This work has started and will be reported on in a future paper.

\begin{figure}[!htb]
    \centering
\includegraphics[width=.55\textwidth,trim=0 0 0 0.56cm, clip]{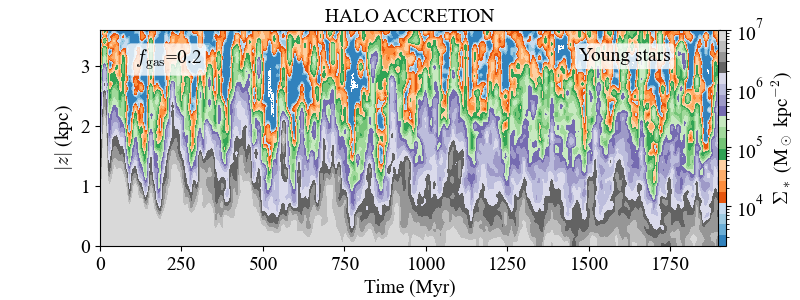}
\includegraphics[width=.35\textwidth,trim=0 0 0 1.1cm, clip]{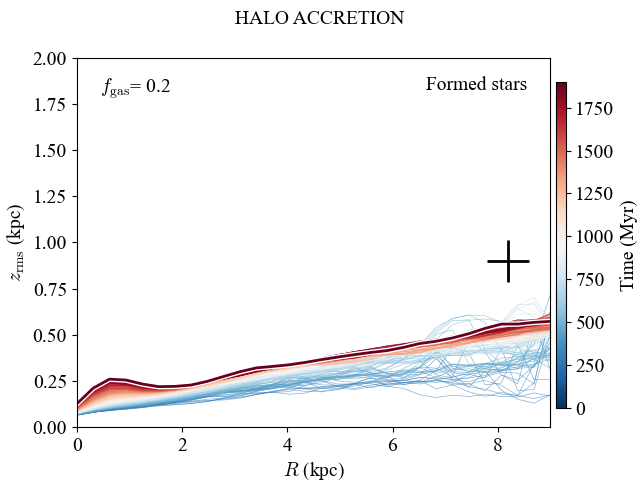}
\includegraphics[width=.55\textwidth,trim=0 0 0 0.56cm, clip]{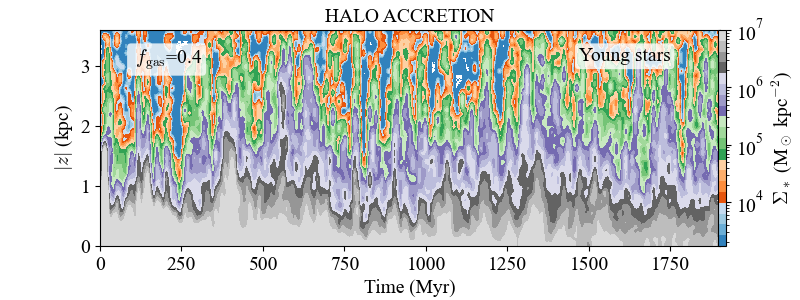}
\includegraphics[width=.35\textwidth,trim=0 0 0 1.1cm, clip]{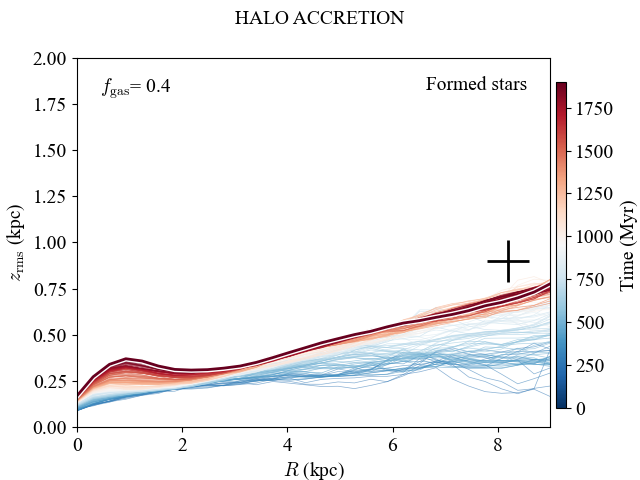}
\includegraphics[width=.55\textwidth,trim=0 0 0 0.56cm, clip]{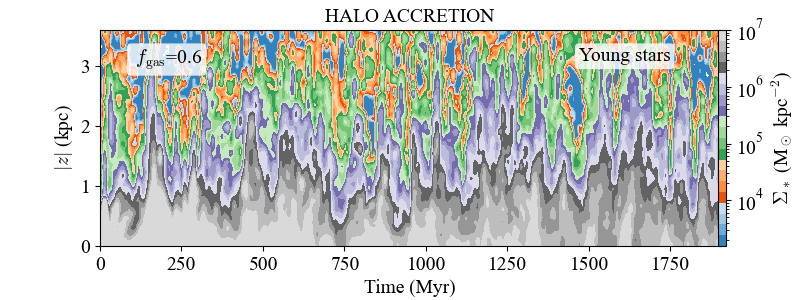}
\includegraphics[width=.35\textwidth,trim=0 0 0 1.1cm, clip]{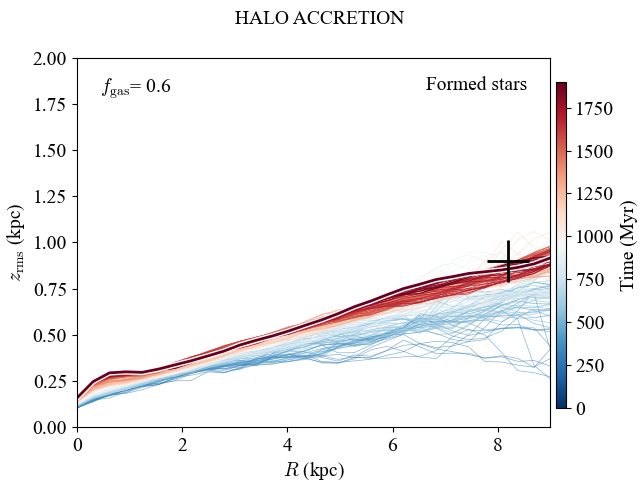}
\includegraphics[width=.55\textwidth,trim=0 0 0 0.56cm, clip]{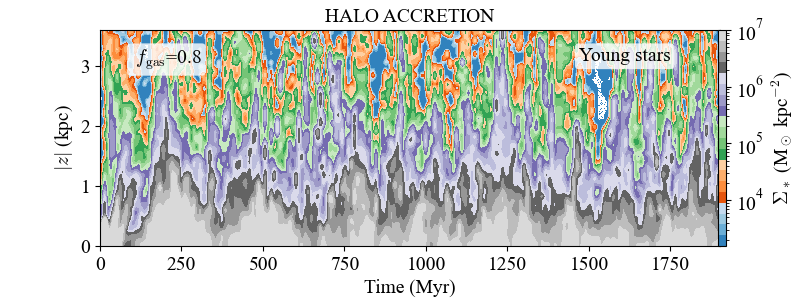}
\includegraphics[width=.35\textwidth,trim=0 0 0 1.1cm, clip]{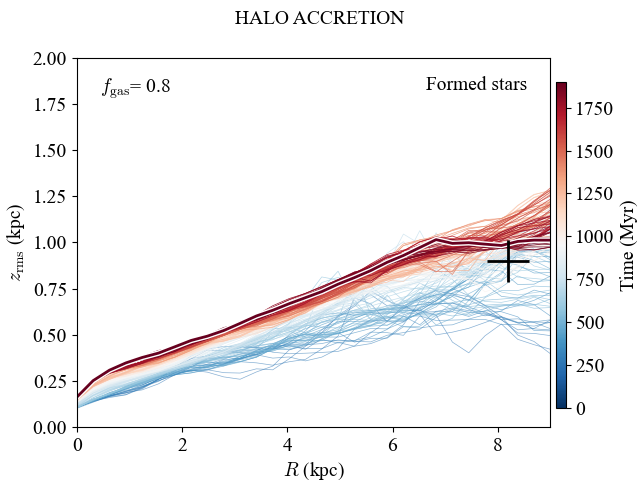}
\includegraphics[width=.55\textwidth,trim=0 0 0 0.56cm, clip]{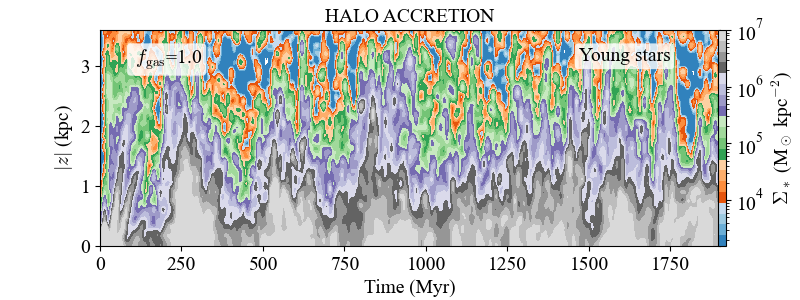}
\includegraphics[width=.35\textwidth,trim=0 0 0 1.1cm, clip]{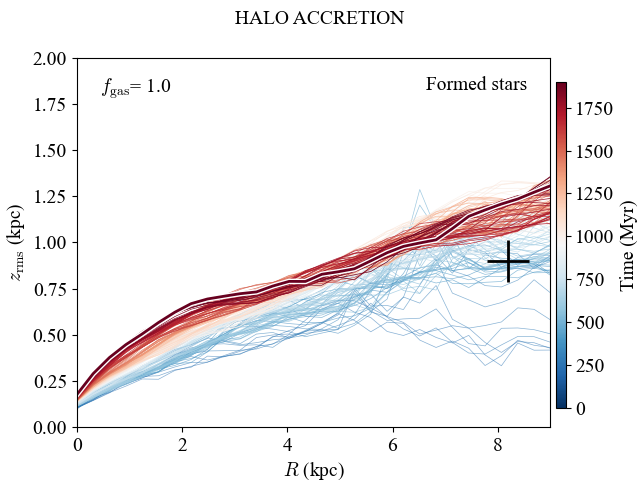}
\caption{(Left) Where new stars are born in time as a function of vertical height $\vert z\vert.$ The left-hand side (LHS) panels look very similar for all $f_{\rm gas}$, quite unlike the RHS panels.
(Right) The rms $z$-scale height for newborn stars as a function of radius $R$ (in kpc) for each value of $f_{\rm gas}$ for the halo accretion cases. In all plots, the colour coding indicates the time evolution, as indicated. The bottom 10 tracks (100 Myr apart) are removed because of large intrinsic scatter; this is a consequence of small number statistics and early populations being born out of dynamical equilibrium.} The scale height of the $\alpha$-rich thick disc today at $R=8$ kpc are indicated by the blue star \citep{bla16}. Note that the vertical scale has a fourfold stretch compared to the horizontal axis. The gas-rich models are able to reproduce the thickness of the $\alpha$-rich disc within $1.5-2$ Gyr.
   \label{f:thicken3}
\end{figure}

\subsection{The case against thick discs ($\vert z\vert \approx 1$ kpc) from in-situ star formation}
\label{s:thickdisc2}

In Secs.~\ref{s:thickdisc1} and \ref{s:internal}, we presented evidence for disc heating being a gradual accumulation of energy from baryon sloshing within the dark matter potential. Star formation feedback and turbulence are implicated in the thickening of discs, but {\it not} from stars being born {\it in situ} far from the cold molecular disc. This idea first arose from non-physical, sticky particle simulations where there is no equation of state, and therefore no pressure balance, or even feedback from star formation processes to support the gas.
In Fig.~\ref{f:thicken3}, most stars in dynamical equilibrium are born close to the plane within the scale height of the cold gas. The remaining 20$-$30\% of stars are injected under non-equilibrium conditions. The LHS figures are similar in their temporal behaviour for all $f_{\rm gas}$, {\it unlike} the RHS figures. In the RHS panel, the disc heating rate for formed stars in time increases rapidly with $f_{\rm gas}$, all starting from a low base set by the maximum dispersion allowed to turbulent-pressure supported molecular gas \citep[$\leq 20-30$ km s$^{-1}$;][]{jef22}. This is because the Brownian motion arising from baryon sloshing heats the stellar disc essentially isotropically, and the amplitude of this motion is a strong function of $f_{\rm gas}$. Eventually, as the younger stellar component settles, its heating rate emerges from the noise and matches the pre-existing population. This is evident by matching the different curves one for one in Fig.~\ref{f:thicken1}, starting with the oldest tracks and moving backwards in time.

In summary, although star formation and turbulence are involved, our mechanism is very different. The energy to thicken the stellar disc is originally from star formation, and then converted to bulk kinetic motion of the baryonic gravitational potential with respect to an inertial frame (approximated by the dark matter frame), which is subsequently diffused into the thermal kinetic energy of the stars. 

\section{Looking forward}
\label{s:summ}

The formation and evolution of disc galaxies is recognized as a difficult problem that is far from resolution. Once again, observations lead theory with the remarkable revelation (Sec.~\ref{s:intro}) of the early onset of discs from JWST and ALMA observations, in particular. However, more observational progress is needed: there is a distinct lack of galaxies with both high-quality JWST {\it and} ALMA data. 

Here we do not consider the early-onset problem. We focus on the simple fact that most discs must have gone through a gas-rich phase given that half of all Milky Way stars were born before $z\sim 1$ when the Universe was less than a third of its current age \citep{fre02d,elb18}. Very little is known about the evolution of gas-rich discs, and {\tt Nexus} was developed with the specific goal of shedding light on this important epoch. What little we do know has come from cosmological simulations at much lower resolution. Here we demonstrate the benefits of increased resolution by revealing new physical processes in these early discs.

In our first {\tt Nexus} paper \citep{bla24}, we showed how bulges, bars and spiral arms emerge even in turbulent, gas-rich environments, consistent with the remarkable new JWST results \citep{guo23,leconte24}. In the present work, we focus on dynamical processes that unbind the disc and cause it to slosh around. The sloshing leads to a thickened stellar disc with properties remarkably consistent with our Galactic $\alpha$-enriched thick disc, in particular, the prediction of vertical gradients in age and metallicity, with only a weak to no radial gradient. These are complex non-linear processes that can only be addressed through numerical simulations presently. Analytic models that describe the evolution of fast and slow fluctuations are under way but at present they have been applied to only a few restricted cases \citep{ham24,bin25}.

There is little doubt that gas-dominated discs were abundant at early times. This has been evident for the past two decades, starting with early molecular gas observations and ground-based integral field studies of ionized gas 
\citep{chap04,gen06h,forster06,shapiro08,swinbank12}.
The most obvious manifestations of turbulent galaxies at high redshift are the observed high gas fractions and large gas clouds \citep[e.g.][]{fen21}, elevated star formation rates and broadened gas kinematics seen in ALMA and JWST surveys (Sec.~\ref{s:intro}). The enhanced star-forming discs at high redshift exhibit larger velocity dispersions than for local discs, regardless of the emission diagnostic used \citep{ejdet22}, and presumably these reflect enhanced levels of turbulent energy. 

A strong prediction of our work is that early discs undergo `baryon sloshing’ with a spatial and kinematic amplitude that gets larger with increasing $f_{\rm gas}$.
Can we find direct evidence of this process? What are the observable manifestations of the predicted sloshing action?
There are already tantalizing clues. \cite{ven20} has shown that quasars display $\sim 20$ km s$^{-1}$ velocity offsets and $\sim$1 kpc spatial offsets in projection with respect to their accompanying gas-rich discs. These amplitudes are consistent with our high $f_{\rm gas}$ models. From JWST imaging, \cite{leb24} find that galaxy disc asymmetries dominate beyond $z\sim 1$ becoming less prevalent at lower redshift; in addition, this behaviour is seen in the absence of companions and in all environments. These recent results may have other explanations but they are entirely consistent with our expectations.

In a companion paper \citep{bla25b}, we explore both morphological and kinematic signatures of baryon sloshing. We highlight specific signatures in low-order ($m=1,2,3$) Fourier components and show that these effects may already be evident in high-redshift ALMA observations \citep[e.g.][]{tsu21,tsu24}.
We focus our study on both the morphology and kinematics of the starlight and the cool gas for high-redshift galaxies up to $z\sim 4$. There are now dozens of galaxies that have been mapped by ALMA in cool gas out to this redshift, as discussed in Sec.~\ref{s:intro}. The notion of stellar kinematics beyond $z\sim 1$ may seem fanciful, except that at least two sources now have (sufficiently) resolved stellar velocity fields in the range $z=3-4$ \citep{eug24,perez25}, with more to come. This was made possible by the remarkable sensitivity of the JWST NIRSpec integral field spectrograph. These are direct line-of-sight observations unaided by gravitational lensing. Reconstructed velocity fields from lensing are also now possible in the same redshift window.
Consistent with our models, these stellar discs are relatively young ($300-700$ Myr) and thus detectable in Balmer absorption lines. For both discs, the star formation rates are of order 10 M$_\odot$ yr$^{-1}$ and considered `quiescent' for such an early epoch, but these rates were an order of magnitude higher at even earlier times.

In future work, already under way, we will look at how merger activity affects our results during the gas-rich phase. It will be important to separate and quantify the respective roles of baryon sloshing and merger activity.
We will evolve the {\tt Nexus} models over the lifetime of the simulated Milky Way-like galaxies. What are the likely chemodynamic signatures that survive? We conjecture that the different baryon components will have different heating signatures, induced by the sloshing, compared to other scenarios. While such orbits can arise from past mergers \citep[e.g.][]{gal19,ama20}, there may be chemodynamic signatures that separate these two distinct processes. If most galaxies are ultimately found to have chemically bimodal discs, given that most discs go through a gas-rich phase in the early Universe, the sloshing mechanism is arguably the most natural process to explain their prevalence.

We are upgrading {\tt Nexus} with more complex gas processes, including MHD and cosmic ray heating, AGN heating, and molecule and dust formation primarily. This will allow our models to be compared to a wider class of objects at different epochs, with varying properties (e.g. total mass, star formation rate) observed across the full electromagnetic spectrum. These are important steps in understanding the full narrative that takes us from early disc systems to modern-day Milky Way counterparts.


\section{Acknowledgments}

JBH acknowledges a Universit\'{e} Paris Science et Lettres (PSL) professorial fellowship that partially supported the writing stage of this work. He is indebted to the graduate students at Paris and Meudon, for their engagement in the graduate course and for inspiring conversations on early disc formation.

JBH is indebted to Merton College and the Beecroft building for providing a stimulating research environment.
We are particularly indebted to James Binney for an advanced copy of his new book {\it Stellar Dynamics} (2025), and for its innovative analysis of dynamical fluctuations that was pivotal for this work. We also thank him for providing the code for the calculation in Sec. 4.2. Moreover, we received valuable feedback from seminars delivered at Marseille, Paris, Shanghai, UCL, Cambridge and Oxford. Important questions were raised by Lia Athanassoula, Chris Hamilton, Francoise Combes, Bruce Elmegreen, Walter Dehnen, Wyn Evans and Vasily Belokurov, and we have sought to answer them here. We are grateful to Eugene Vasiliev and Romain Teyssier for continued assistance and adaptions with \agama\ and \ramses\ respectively. We are also indebted to a thorough and perceptive referee.

TTG acknowledges financial support from the Australian Research Council (ARC) through Australian Laureate Fellowships awarded to JBH (FL140100278) and TRB (FL220100117), and for partial funding through the James Arthur Pollock memorial fund awarded to the School of Physics, University of Sydney.
OA acknowledges support from the Knut and Alice Wallenberg Foundation, the Swedish Research Council (grant 2019-04659) and the Swedish National Space Agency (SNSA Dnr 2023-00164). CF acknowledges funding provided by the Australian Research Council (Discovery Project DP230102280 and DP250101526), and the Australia-Germany Joint Research Cooperation Scheme (UA-DAAD).

The computations and data storage were enabled by two facilities: (i) the National Computing Infrastructure (NCI) Adapter Scheme, provided by NCI Australia, an NCRIS capability supported by the Australian Government; and (ii) LUNARC, the Centre for Scientific and Technical Computing at Lund University (resource allocations LU 2023/2-39 and LU 2023/12-6). We further acknowledge high-performance computing resources provided by the Leibniz Rechenzentrum and the Gauss Centre for Supercomputing (grants~pr32lo, pr48pi, pn76ga and GCS Large-scale project~10391), the Australian National Computational Infrastructure (grant~ek9) and the Pawsey Supercomputing Centre (project~pawsey0810) in the framework of the National Computational Merit Allocation Scheme and the ANU Merit Allocation Scheme.



\bigskip
\appendix
\section{Radial and kinematic signatures of baryon sloshing}
\label{s:signatures}

\begin{figure}[!htb]
    \centering
\includegraphics[width=\textwidth]{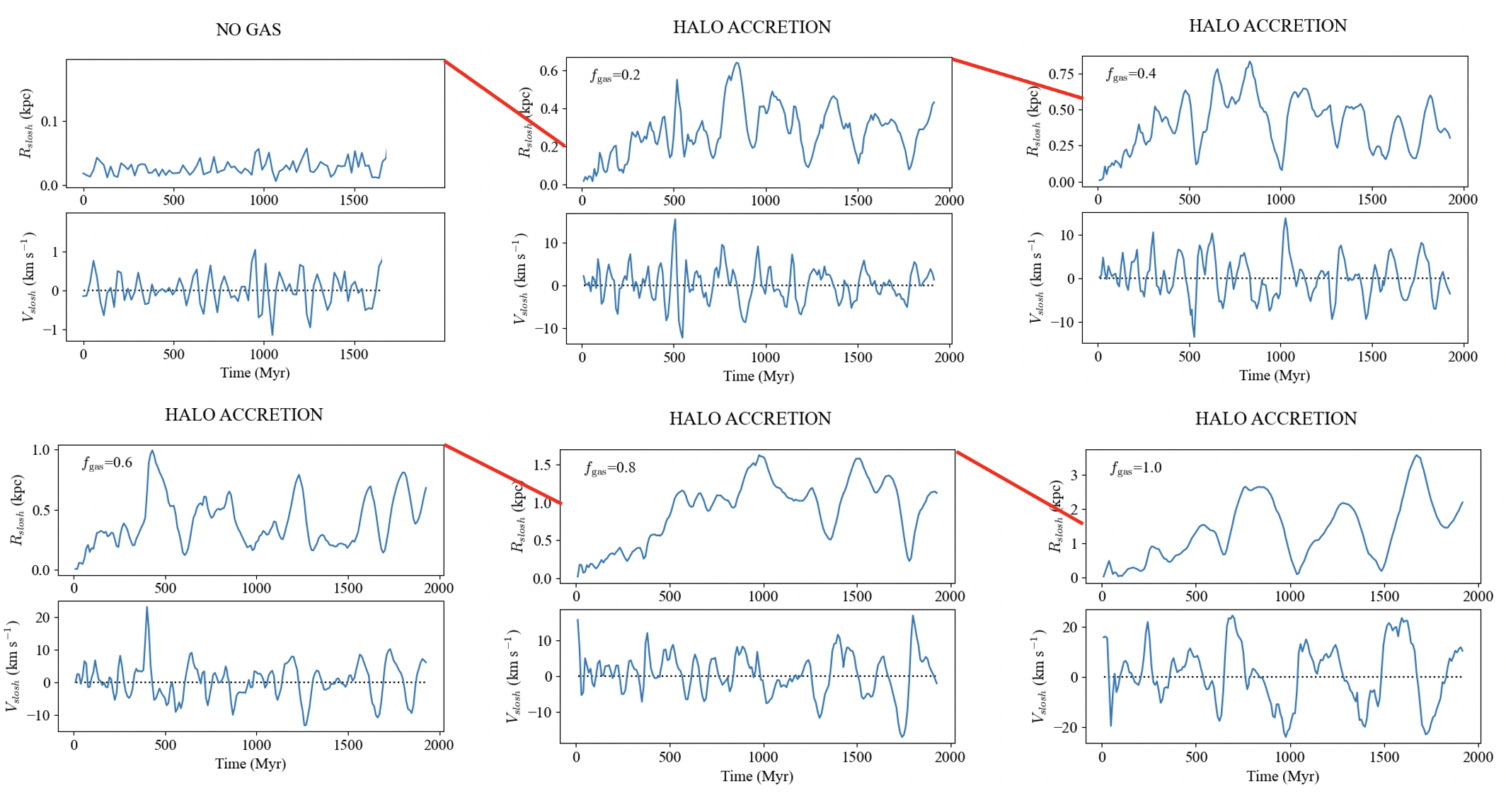}
\caption{Baryon sloshing presented as the change in radius $R_{\rm slosh}=\sqrt{x_{\rm slosh}^2+y_{\rm slosh}^2+z_{\rm slosh}^2}$ with cosmic time (upper panels)} in the reference frame of the total potential for $f_{\rm gas}=0,20,40,60,80,100$\%. The lower panels are the derivative $\dot{R}_{\rm slosh}$ of the upper panels. The red lines indicate the changing physical scales between panels, where the overall trend is increasing with $f_{\rm gas}$. Note also that the sloshing frequency declines with increasing $R$, as expected from Eq.~\ref{e:freq}.
   \label{f:rv_slosh}
\end{figure}

In Sec.~\ref{s:slosh}, we presented initial evidence of the 3D baryon sloshing, with kinematic and radial information compared in Fig.~\ref{f:slosh1} (Right) for a few specific cases. The full gallery of radial and kinematic signatures is presented in Fig.~\ref{f:rv_slosh}. As expected, the radial and kinematic excursion amplitudes increase with $f_{\rm gas}$. Note also the decreasing frequency of oscillation as the baryon centroid wanders further and further from the centre of total gravitational potential. This is a natural consequence of the radial trends in Eq.~\ref{e:freq} (see also Eq.~\ref{e:zfreq}).

\bibliographystyle{aasjournal}

\input{main.bbl}

\end{document}

%% file: defs_custom.tex
\definecolor{orange}{rgb}{1.0,0.5,0.}

\def\MDM{\ifmmode{\>M_{\textnormal{\sc dm}}}\else{$$M_{\textnormal{\sc dm}}}\fi}

\def\XH{\ifmmode{\>X_{\textnormal{\sc h}}} \else{$X_{\textnormal{\sc h}}$}\fi}
\def\nH{\ifmmode{\>n_{\textnormal{\sc h}}} \else{$n_{\textnormal{\sc h}}$}\fi}

\def\maspyr{\ifmmode{\>\textnormal{mas~yr}^{-1}}\else{mas~yr$^{-1}$}\fi}

\def\mG{\ifmmode{\>\mu\mathrm{G}}\else{$\mu$G}\fi}
\def\erg{\ifmmode{\> {\rm erg}}\else{erg}\fi}
\def\keV{\ifmmode{\> {\rm keV}}\else{keV}\fi}

\def\deg{\ifmmode{\>^{\circ}}\else{$^{\circ}$}\fi}
\def\onedeg{\ifmmode{\>1^{\circ}}\else{$1^{\circ}$}\fi}

\def\xvir{\ifmmode{\>\!x_{vir}}\else{$x_{vir}$}\fi}
\def\Mvir{\ifmmode{\>\!M_{vir} }\else{$M_{vir} $}\fi}
\def\rvir{\ifmmode{\>\!r_{vir}}\else{$r_{vir}$}\fi}
\def\vvir{\ifmmode{\>\!v_{vir}}\else{$v_{vir}$}\fi}
\def\Vvir{\ifmmode{\>\!V_{vir} }\else{$V_{vir} $}\fi}

\def\tratio{\ifmmode{\>\tau}\else{$\tau$}\fi}

\def\rms{\ifmmode{\>r_{\textnormal{\sc ms}}}\else{$r_{\textnormal{\sc ms}}$}\fi}

\def\Mpc{\ifmmode{\>\!{\rm Mpc}} \else{Mpc}\fi}
\def\kpc{\ifmmode{\>\!{\rm kpc}} \else{kpc}\fi}
\def\pkpc{\ifmmode{\>\!{\rm kpc}^{-1}} \else{kpc$^{-1}$}\fi}
\def\pc{\ifmmode{\>\!{\rm pc}} \else{pc}\fi}

\def\Gyr{\ifmmode{\>\!{\rm Gyr}} \else{Gyr}\fi}
\def\Myr{\ifmmode{\>\!{\rm Myr}} \else{Myr}\fi}
\def\yr{\ifmmode{\>\!{\rm yr}} \else{yr}\fi}
\def\pyr{\ifmmode{\>\!{\rm yr}^{-1}}\else{yr$^{-1}$} \fi}
\def\s{\ifmmode{\>\!{\rm s}}\else{s}\fi}
\def\ps{\ifmmode{\>\!{\rm s}^{-1}}\else{s$^{-1}$}\fi}
\def\Hz{\ifmmode{\>\!{\rm Hz}}\else{Hz}\fi}

\def\kms{\ifmmode{\>\!{\rm km\,s}^{-1}}\else{km~s$^{-1}$}\fi}

\def\K{\ifmmode{\>\!{\rm K}}\else{K}\fi}

\def\sr{\ifmmode{\>\!{\rm sr}}\else{sr}\fi}
\def\psr{\ifmmode{\>\!{\rm sr}^{-1}}\else{sr$^{-1}$}\fi}
\def\arcs{\ifmmode{\>\!{\rm arcsec}}\else{arcsec}\fi}
\def\parcs{\ifmmode{\>\!{\rm arcsec}^{-1}}\else{arcsec${-1}$}\fi}
\def\parcss{\ifmmode{\>\!{\rm arcsec}^{-2}}\else{arcsec${-2}$}\fi}

\def\cm{\ifmmode{\>\!{\rm cm}}\else{cm}\fi}
\def\cc{\ifmmode{\>\!{\rm cm}^{3}}\else{cm$^{3}$}\fi}
\def\sqc{\ifmmode{\>\!{\rm cm}^{2}}\else{cm$^{2}$}\fi}
\def\pcc{\ifmmode{\>\!{\rm cm}^{-3}}\else{cm$^{-3}$}\fi}
\def\psc{\ifmmode{\>\!{\rm cm}^{-2}}\else{cm$^{-2}$}\fi}

\def\g{\ifmmode{\>\!{\rm g}}\else{g}\fi}
\def\Msun{\ifmmode{\>\!{\rm M}_{\odot}}\else{M$_{\odot}$}\fi}
\def\hMsun{\ifmmode{\> h^{-1}{\rm M}_{\odot}}\else{$h^{-1}$M$_{\odot}$}\fi}

\def\Zsun{\ifmmode{\>\!{\rm Z}_{\odot}}\else{Z$_{\odot}$}\fi}

\def\Lsun{\ifmmode{\>\!{\rm L}_{\odot}}\else{L$_{\odot}$}\fi}

\def\rayl{\ifmmode{\>\!{\rm R}}\else{R}\fi}
\def\mR{\ifmmode{\>\!{\rm mR}}\else{mR}\fi}

\def\lya{\ifmmode{\>\!{\rm Ly}\alpha}\else{Ly$\alpha$}\fi}

\def\Ha{\ifmmode{\>\!{\rm H}\alpha}\else{H$\alpha$}\fi}
\def\Hb{\ifmmode{\>\!{\rm H}\beta}\else{H$\beta$}\fi}

\def\HI{\ifmmode{\> \textnormal{\ion{H}{1}}} \else{\ion{H}{1}}\fi}
\def\HII{\ifmmode{\> \textnormal{\ion{H}{2}}} \else{\ion{H}{2}}\fi}
\def\CIV{\ifmmode{\> \textnormal{\ion{C}{4}}} \else{\ion{C}{4}}\fi}
\def\SiIV{\ifmmode{\> \textnormal{\ion{S}{4}}} \else{\ion{Si}{4}}\fi}

\def\NH{\ifmmode{\> {\rm N}_{\rm H}} \else{N$_{\rm H}$}\fi}
\def\Ng{\ifmmode{\> {\rm N}_{\rm gas}} \else{N$_{\rm gas}$}\fi}
\def\NHI{\ifmmode{\> {\rm N}_{\HI}} \else{N$_{\HI}$}\fi}
\def\MHI{\ifmmode{\> {\rm M}_{ \HI}} \else{M$_{\HI}$}\fi}

\def\mua{\ifmmode{\>\mu_{ \textnormal{\Ha}}}\else{$\mu_{ \textnormal{\Ha}}$}\fi}
\def\alphabha{\ifmmode{\>\alpha_{B}^{(\textnormal{\Ha})}}\else{$\alpha_{B}^{(\textnormal{\Ha})}$}\fi}

%% file: authors.tex

\author[0000-0001-7516-4016]{Joss Bland-Hawthorn}
\affiliation{Sydney Institute for Astronomy, School of Physics, A28, The University of Sydney, NSW 2006, Australia}
\affiliation{Centre of Excellence for All-Sky Astrophysics in Three Dimensions (ASTRO 3D), Australia}

\author[0000-0002-1081-883X]{Thor Tepper-Garcia}
\affiliation{Sydney Institute for Astronomy, School of Physics, A28, The University of Sydney, NSW 2006, Australia}
\affiliation{Centre of Excellence for All-Sky Astrophysics in Three Dimensions (ASTRO 3D), Australia}

\author[0000-0002-4287-1088]{Oscar Agertz}
\affiliation{Lund Observatory, Division of Astrophysics, Department of Physics, Lund University, Box 43, SE-221 00 Lund, Sweden}

\author[0000-0002-0706-2306]{Christoph Federrath}
\affiliation{Research School of Astronomy and Astrophysics, Australian National University, Canberra, ACT 2611, Australia}
\affiliation{Centre of Excellence for All-Sky Astrophysics in Three Dimensions (ASTRO 3D), Australia}

\author[0000-0003-1623-6643]{Misha Haywood}
\affiliation{GEPI, Observatoire de Paris, CNRS, Université Paris Diderot, 5 place Jules Janssen, 92190 Meudon, France}

\author[0000-0002-5213-4807]{Paola di Matteo}
\affiliation{GEPI, Observatoire de Paris, CNRS, Université Paris Diderot, 5 place Jules Janssen, 92190 Meudon, France}

\author[0000-0001-5222-4661]{Timothy R Bedding}
\affiliation{Sydney Institute for Astronomy, School of Physics, A28, The University of Sydney, NSW 2006, Australia}

\author[0000-0002-1499-6377]{Takafumi Tsukui}
\affiliation{Research School of Astronomy and Astrophysics, Australian National University, Canberra, ACT 2611, Australia}

\author[0000-0003-1657-7878]{Emily Wisnioski}
\affiliation{Research School of Astronomy and Astrophysics, Australian National University, Canberra, ACT 2611, Australia}

\author[0000-0001-5082-6693]{Melissa Ness}
\affiliation{Research School of Astronomy and Astrophysics, Australian National University, Canberra, ACT 2611, Australia}

\author[0000-0001-6280-1207]{Ken Freeman}
\affiliation{Research School of Astronomy and Astrophysics, Australian National University, Canberra, ACT 2611, Australia}
